%% file: CODATA_NP_01.tex
\documentclass[aps,prd,preprintnumbers,superscriptaddress,nofootinbib,floatfix,twocolumn,notitlepage]{revtex4-2}
\usepackage{graphicx}  % needed for figures
\usepackage{dcolumn}   % needed for some tables
\usepackage{bm}        % for math
\usepackage{epsfig,amsmath,amssymb,verbatim,mathrsfs,array,layout,textcomp,amssymb,latexsym,slashed}
\usepackage{xcolor}
\usepackage[colorlinks=true,citecolor=blue,urlcolor=blue,linktocpage=true,
linkcolor=blue]{hyperref}
\usepackage[utf8]{inputenc}
\usepackage{multirow}
\usepackage{booktabs} %for \toprule
\usepackage{siunitx} %for \num
\usepackage{tabularx} %for tabularx

\newcommand{\cL}{\mathcal{L}}

\newcommand{\cO}{\mathcal{O}}

\newcommand{\ie}{\textit{i.e.}}

\newcommand{\keV}{{\rm keV}}
\newcommand{\MeV}{{\rm MeV}}

\newcommand{\nuh}{\nu_{\rm H}}
\newcommand{\nud}{\nu_{\rm D}}
\newcommand{\dH}{\delta_{\rm H}}
\newcommand{\dD}{\delta_{\rm D}}
\newcommand{\dHDplus}{\delta_{\rm HD^+}}
\newcommand{\dHefourpbar}{\delta_{\rm \bar{p}^4He}}
\newcommand{\dHethreepbar}{\delta_{\rm \bar{p}^3He}}

\newcommand{\SM}{{\rm SM}}
\newcommand{\alphaEM}{\alpha}
\newcommand{\Rinf}{R_{\infty}}

\newcommand{\NP}{{\rm NP}}
\newcommand{\alphaNP}{\alpha_{\phi}}
\newcommand{\mNP}{m_{\phi}}

\newcommand{\be}{\begin{equation}}
\newcommand{\ee}{\end{equation}}
\newcommand{\bea}{\begin{eqnarray}}
\newcommand{\eea}{\end{eqnarray}}

\newcommand{\hdplus}{{$\text{HD}^+ $}}
\newcommand{\hefourpbar}{{$\bar{{p}}{}^4\text{He}$}}
\newcommand{\hethreepbar}{{$\bar{{p}}{}^3\text{He}$}}
\newcommand{\hepbar}{{$\bar{{p}}\text{He}\ $}}

\definecolor{light_blue}{rgb}{0.15, 0.35, 0.9}
\newcommand{\OLD}[1]{\textcolor{black}{#1}}

\begin{document}

\title{Self-consistent extraction of spectroscopic bounds on light new physics}
\preprint{LAPTH-063/22}
\preprint{CERN-TH-2022-158}
\preprint{KEK-TH-2454}

%%%%%%%%%%%%%%%%%%%%%%%%%%%%%%%%%%%%%%%%%%%%%%%%%%%%%%%%%%%%%%%%%%%%%%
\author{C\'edric Delaunay}
\email{cedric.delaunay@lapth.cnrs.fr}
\affiliation{Laboratoire d'Annecy-le-Vieux de Physique Th\'eorique, CNRS -- USMB, BP 110 Annecy-le-Vieux, F-74941 Annecy, France}
\affiliation{Theoretical Physics Department, CERN,
Esplanade des Particules 1, Geneva CH-1211, Switzerland}

\author{Jean-Philippe Karr}
\email{karr@lkb.upmc.fr}
\affiliation{Laboratoire Kastler Brossel, Sorbonne Universit\'e, CNRS, ENS-Universit\'e PSL, Coll\`ege de France, 4 place Jussieu, F-75005 Paris, France}
\affiliation{Universit\'e d'Evry-Val d'Essonne, Universit\'e Paris-Saclay, Boulevard Fran\c{c}ois Mitterrand, F-91000 Evry, France}

\author{Teppei Kitahara}
\email{teppeik@kmi.nagoya-u.ac.jp}
\affiliation{Institute for Advanced Research \& Kobayashi-Maskawa Institute for the Origin of Particles and the Universe, 
Nagoya University,  Nagoya 464--8602, Japan}
\affiliation{KEK Theory Center, IPNS, KEK, Tsukuba  305--0801, Japan}
\affiliation{CAS Key Laboratory of Theoretical Physics, Institute of Theoretical Physics, Chinese Academy of Sciences, Beijing 100190, China}

\author{Jeroen C. J. Koelemeij}
\email{j.c.j.koelemeij@vu.nl}
\affiliation{LaserLaB, Department of Physics and Astronomy, Vrije Universiteit Amsterdam, De Boelelaan
1081, 1081 HV Amsterdam, The Netherlands}

\author{Yotam Soreq}
\email{soreqy@physics.technion.ac.il}
\affiliation{Physics Department, Technion -- Israel Institute of Technology, Haifa 3200003, Israel}

\author{Jure Zupan}
\email{zupanje@ucmail.uc.edu}
\affiliation{Department of Physics, University of Cincinnati, Cincinnati, Ohio 45221,USA}
%%%%%%%%%%%%%%%%%%%%%%%%%%%%%%%%%%%%%%%%%%%%%%%%%%%%%%%%%%%%%%%%%%%%%%

%%%%%%%%%%%%%%%%%%%%%%%%%%%%%%%%%%%%%%%%%%%%%%%%%%%%%%%%%%%%%%%%%%%%%%
\begin{abstract}
Fundamental physical constants are determined from a collection of precision  measurements of elementary particles, atoms and molecules. 
This is usually done under the assumption of the Standard Model~(SM) of particle physics. 
Allowing for light new physics~(NP) beyond the SM modifies the extraction of fundamental physical constants. 
Consequently, setting NP bounds using these data, and at the same time assuming the CODATA recommended values for the fundamental physical constants, is not reliable. 
As we show in this Letter, both SM and NP parameters can be simultaneously determined in a consistent way from a global fit.
For light vectors with QED-like couplings, such as the dark photon, we provide a prescription that recovers the degeneracy with the photon in the massless limit, and requires calculations only at leading order in the small new physics couplings.
At present, the data show tensions partially related to the proton charge radius 
determination. 
We show that these can be alleviated by including contributions from a light scalar with flavor non-universal couplings. 
\end{abstract}
%%%%%%%%%%%%%%%%%%%%%%%%%%%%%%%%%%%%%%%%%%%%%%%%%%%%%%%%%%%%%%%%%%%%%%

\maketitle

%%%%%%%%%%%%%%%%%%%%%%%%%%%%%%%%%%%%%%%%%%%%%%%%%%%%%%%%%%%%%%%%%%%%%%
\section{Introduction}
\label{sec:Intro}
%%%%%%%%%%%%%%%%%%%%%%%%%%%%%%%%%%%%%%%%%%%%%%%%%%%%%%%%%%%%%%%%%%%%%%

Precision measurements of atomic and molecular properties play a dual role in fundamental physics. 
On the one hand, assuming the Standard Model~(SM) of particle physics, these are used to determine two of the SM parameters, the fine-structure constant, $\alpha$, and the electron mass, $m_e$ (through the Rydberg constant $\Rinf\equiv \alpha^2m_e c/(2h)$), along with a number of other observables such as the charge radii and relative atomic masses of the proton and deuteron. 
An example is the determination of fundamental physical constants by the Committee on Data of the International Science Council~(CODATA)~\cite{Tiesinga:2021myr}.

On the other hand, precision measurements can be used 
to search for 
new physics~(NP) beyond the SM. 
Such searches have been conducted using measurements of single particle observables~\cite{Hanneke:2008tm,Muong-2:2021ojo,ACME:2018yjb}, atomic systems~\cite{Jaeckel:2010xx,Karshenboim:2010ck,Delaunay:2016brc,Berengut:2017zuo,Frugiuele:2019drl,Frugiuele:2021bic}, and molecular systems~\cite{Salumbides:2013aga,Borkowski:2016zas,Schiller:HDp,Germann:2021koc}, see~\cite{Safronova:2017xyt} for a review. 
The presence of NP would manifest itself as a discrepancy between measurements and theoretical SM predictions. 
The difficulty here is that in many cases the SM predictions depend on the fundamental physics parameters, which in turn were extracted from data by CODATA \textit{under the assumption that the SM is correct, and no NP exists}. 
In general, the presence of NP would affect the extraction of fundamental constants, possibly reducing the claimed sensitivity of NP searches. 
This subtlety is more often than not ignored in the literature. 
 
In this Letter we propose and carry out a self-consistent determination of constraints on light NP models by performing a global fit, simultaneously extracting the SM and NP parameters. 
We go well beyond the previous studies~\cite{Jaeckel:2010xx,Karshenboim:2010ck,Jones:2019qny}, which were performed only on subsets of data. 
We pay special attention to the potentially problematic limit of massless NP.  
The challenge is that the SM predictions are calculated to a higher perturbative order than the leading order~(LO) NP contributions, which can then lead to incorrect limiting behaviour for very light NP. 
Below, we provide a prescription, valid to LO in NP parameters, that corrects for such mismatches in the theoretical predictions, and leads to the proper massless NP limit.

The global fit shows several $3\,\sigma$ (3 standard deviations) discrepancies between observables and predictions, assuming the SM. 
These anomalies are well known: 
they correspond to the measurements constituting the proton charge radius puzzle~\cite{Pohl:2010zza,Antognini:2013txn,Karr:2020}, with the addition of new measurements of hydrogen transitions~\cite{Brandt:2021yor,Grinin:2020}. 
Reference~\cite{Brandt:2021yor} showed the tension of their $2S_{1/2}-8D_{5/2}$ measurement with other hydrogen data is relaxed in the presence of an additional Yukawa-like interaction.
Our global analysis, which determines simultaneously both the SM and NP parameters, shows for the first time that all these deviations can be largely accounted for in a single NP model -- a light scalar that couples to gluons, electrons and muons.

%%%%%%%%%%%%%%%%%%%%%%%%%%%%%%%%%%%%%%%%%%%%%%%%%%%%%%%%%%%%%%%%%%%%%%
\section{New Physics Benchmark Models}
\label{sec:NPModels}
%%%%%%%%%%%%%%%%%%%%%%%%%%%%%%%%%%%%%%%%%%%%%%%%%%%%%%%%%%%%%%%%%%%%%%

We focus on minimal extensions of the SM, where either a light scalar boson, $\phi$, or a light vector boson, $\phi_\mu$, is added to the spectrum of SM particles. 
The new light particle is assumed to have parity conserving interactions with the SM electrons and muons, as well as with light quarks, resulting in couplings to neutrons and protons\footnote{Extension to parity non-conserving couplings and additional particles is straightforward.}. 
The interaction Lagrangian is therefore given by 
\begin{align}
    \label{eq:Lint}
    \cL_{\rm int} 
=   \sum_{i=e,\mu,n,p} g_i \overline{\psi}_i (\Gamma\cdot\phi) \psi_i\,,
\end{align}
where $\Gamma\cdot\phi \equiv \phi, \gamma^\mu\phi_\mu$ for spin $s=0, 1$ bosons, respectively. 
Taking the nonrelativistic limit for $\psi_i$, and working at LO in $g_i$, the tree level exchange of $\phi$ or $\phi_\mu$ induces a Yukawa-like nonrelativistic potential,
\begin{align}
    \label{eq:VNP}
    V_{\rm NP}^{ij}(r)
=   (-1)^{s+1}\alphaNP q_iq_j\frac{e^{-\mNP r}}{r}\,, 
\end{align}
between particles $\psi_i$ and $\psi_j$, separated by a distance $r$.
The NP coupling constant, $\alphaNP\equiv |g_eg_p|/(4\pi)>0$, gives the strength of the NP induced potential between electrons and protons. 
The strength of NP interactions between fermions $\psi_i$ and $\psi_j$, relative to the electron--proton one, is given by the product of effective NP couplings, $q_iq_j$, where $q_i\equiv g_i/\sqrt{|g_eg_p|}$. 
In particular, for the electron--proton system the product of effective NP couplings can take the values, $q_eq_p=\pm1$. 
For $q_iq_j>0$ the potential \eqref{eq:VNP} is attractive\,(repulsive) for spin 0\,(1) mediator $\phi$, and vice versa for $q_i q_j<0$.

In the numerical analysis, we consider the following benchmark NP  models: 
\\
\paragraph{Dark photon.} The light NP mediator is a vector boson with couplings to the SM fermions proportional to their electric charges. 
A UV complete realization is 
an additional abelian gauge boson with field strength $F_{\mu\nu}'$, that couples to the SM through the renormalizable kinetic mixing interaction,  $-(\epsilon/2)F'_{\mu\nu}F^{\mu\nu}$~\cite{Holdom:1985ag}, where $F_{\mu\nu}$ is the electromagnetic field strength.
To LO in $\epsilon$ this yields $\alphaNP=\alpha\epsilon^2$ and $q_e=q_\mu=-q_p=-1$, $q_n=0$.
\\
\paragraph{$B-L$ gauge boson.} 
The difference of baryon ($B$) and lepton ($L$) numbers is non-anomalous, and can be gauged without introducing new fermions \cite{Davidson:1978pm,Marshak:1979fm}.
Light $B-L$ gauge boson with gauge coupling $g_{B-L}$ gives rise to the NP potential in \eqref{eq:VNP} with $\alphaNP=g_{B-L}^2/(4\pi)$. 
The charges $q_e=q_\mu=-q_p=-q_n=-1$ coincide with the dark photon ones, except for neutron.  
Comparison of $B-L$ and  dark photon bounds illustrates
the importance of performing spectroscopy of different isotopes of the same species, such as hydrogen and deuterium.
\\ 
\paragraph{Scalar Higgs portal.}
A light scalar  mixing with the Higgs boson~\cite{Patt:2006fw,OConnell:2006rsp} 
inherits the SM Yukawa structure, giving 
$\alphaNP= \sin^2\theta \, m_e \kappa_p m_p/(4\pi v^2)\lesssim 1.8\times 10^{-10}$ where $v\simeq 246\,$GeV is the SM Higgs vacuum expectation value, and $\theta$ the scalar mixing angle. 
The effective leptonic ($\ell=e,\mu$) charges are $q_\ell=m_\ell/\sqrt{ m_e \kappa_p m_p}$, while the effective nucleon charges ($N=p,n$) are given by $q_N=\kappa_N m_N/\sqrt{ m_e \kappa_p m_p}$ with $\kappa_p\simeq 0.306(14)$ and $\kappa_n\simeq 0.308(14)$~\cite{Shifman:1978zn,Belanger:2008sj,Junnarkar:2013ac,Belanger:2013oya,Bishara:2017pfq,Bishara:2017nnn} (see also Sec.~\ref{sec:further:NP}). 
Since couplings to muons and nucleons are enhanced by $q_\mu/q_e=m_\mu/m_e\simeq 200$ and $g_N/q_e=m_N/m_e\simeq 2\times 10^3$, respectively, this NP benchmark highlights the relevance of muonic atom and molecular spectroscopy.
\\
\paragraph{Hadrophilic scalar.}
A scalar with $q_\ell=0$ and $q_N=\kappa_N m_N/\sqrt{m_e \kappa_p m_p}$, {\it i.e.}, with vanishing couplings to leptons, highlights the importance of molecular hydrogen ion spectroscopy as a probe of internuclear interactions~\cite{Salumbides:2013aga,Germann:2021koc}. 
For expedience we take $g_N$ to be the same as for the Higgs portal, but this could be relaxed in general.
\\  
\paragraph{Up-lepto-darko-philic~(ULD) scalar.}
In order to evade strong bounds from $K^+\to \pi^++X_{\rm inv}$ searches, where $X_{\rm inv}$ are invisible particles that escape the detector, see Sec.~\ref{sec:result}, we adopt a particular version of a light scalar benchmark. The ULD scalar has enhanced couplings to leptons, $q_\ell=m_\ell/\sqrt{m_e\kappa_p' m_p}$, and reduced couplings to nucleons (due to couplings to only the up quark),  $q_N=\kappa_N'm_N/\sqrt{m_e\kappa_p'm_p}$, with  $\kappa_p'\simeq0.018(5)$
 and $\kappa_n'\simeq 0.016(5)$, and $\alpha_\phi=k^2 m_e\kappa_p'm_p/(4\pi v^2)$, with  $k$ a dimensionless parameter controlling the overall strength of interactions, which is varied in the fit. 
The $\phi$ is assumed to predominantly decay to invisible states, possibly related to the dark matter, which evades constraints from beam dump experiments. 
See Sec.~\ref{sec:further:NP} in the supplemental material for further details, including results for an additional NP benchmark model-- the scalar photon.

%%%%%%%%%%%%%%%%%%%%%%%%%%%%%%%%%%%%%%%%%%%%%%%%%%%%%%%%%%%%%%%%%%%%%%
\section{Datasets} 
\label{sec:data}
%%%%%%%%%%%%%%%%%%%%%%%%%%%%%%%%%%%%%%%%%%%%%%%%%%%%%%%%%%%%%%%%%%%%%%

The adjustment of parameters, {\it i.e.}, the fitting procedure, presented in this work has been carried out using two different datasets, CODATA18 and DATA22. 
The CODATA18 dataset consists of data that was used in the latest CODATA adjustment in Ref.~\cite{Tiesinga:2021myr}, but restricted only to the subset most relevant for constraining NP.  
This subset contains observables related to the determination of the Rydberg constant $R_\infty$,  the proton and deuteron radii, $r_p$ and $r_d$ respectively, 
the fine-structure constant  $\alphaEM$, and the relative atomic masses of the electron, proton, and deuteron: $A_{\rm r}(e)$, $A_{\rm r}(p)$ and $A_{\rm r}(d)$, respectively.
The inputs are listed in Tables~\ref{tab:inputsA}, \ref{tab:inputsC}, and include theory uncertainties in Table~\ref{tab:HDuncert}.  
The other observables and parameters included in the CODATA~2018 adjustment are very weakly correlated  with the selected data, and can be neglected for our purpose. 

The DATA22 dataset combines the updated CODATA18 inputs with the additional data  that improve the overall sensitivity to NP (see Table~\ref{tab:inputsNEW} and \ref{tab:3buncert}). 
In particular, we include 
the measurements of transition frequencies in simple molecular or molecule-like systems, the hydrogen deuteride molecular ion (\hdplus)~\cite{Alighanbari:2020,Patra:2020,Kortunov:2021rfe}, and the antiprotonic helium atom (\hethreepbar~and  \hefourpbar)~\cite{Hori:2011,Hori:2016}.  
These have an enhanced sensitivity to the NP models with mediators that have large couplings to quarks (and thus nuclei).  
The three benchmark models of this type are the Higgs portal, hadrophilic  and ULD scalars, cf. Sec~\ref{sec:NPModels}.  

The CODATA18 dataset is used as a reference point to verify the implementation of the inputs and the adjustment procedure, while DATA22 is used to obtain our nominal results.
The full list of data in the two datasets, as well as further discussion of the importance of including certain observables when constraining NP, is given in Supplemental Material, Sections.~\ref{sec:CODATA2018dataset} and \ref{sec:DATA22dataset}.

%%%%%%%%%%%%%%%%%%%%%%%%%%%%%%%%%%%%%%%%%%%%%%%%%%%%%%%%%%%%%%%%%%%%%%%%%%
\section{Least-squares Adjustment with new physics}
\label{sec:fit}
%%%%%%%%%%%%%%%%%%%%%%%%%%%%%%%%%%%%%%%%%%%%%%%%%%%%%%%%%%%%%%%%%%%%%%%%%%%

The experimental data are compared to the theoretical predictions with NP following the linearized least-squares procedure~\cite{Mohr:2000ie}.
The theoretical prediction for an observable $\cO$ takes the form,  
\begin{align}
    \label{eq:th:predict}
    \cO
=   \cO_{\rm SM}(g_{\rm SM}) 
    + \cO_{\rm NP}(g_{\rm SM},\alphaNP,\mNP)
    +\delta\cO_{\rm th} \,,
\end{align}
where $\cO_{\rm SM}$ is the state of the art SM prediction, and depends on the SM parameters $g_{\rm SM}=\{R_\infty,r_p,r_d,\alpha,A_{\rm r}(e),A_{\rm r}(p),
A_{\rm r}(d)\}$, while 
the NP contribution $\cO_{\rm NP}$ depends in addition on $\alphaNP$ and $\mNP$.   
The theoretical uncertainties are included as in Ref.~\cite{Tiesinga:2021myr}, by adding a normally distributed  variable $\delta\cO_{\rm th}$ with zero mean and standard deviation equal to the estimated uncertainty of the theoretical expression. 
The $\delta\cO_{\rm th}$'s are treated as yet another set of input data and varied in the fit, along with $g_{\rm SM}$, $\alphaNP$, and $\mNP$, in order to minimize the $\chi^2$ function constructed from the input data and theory predictions (see also Sec.~\ref{sec:app:fit}).

The SM theoretical predictions for atomic transition frequencies, the electron anomalous magnetic moment, and bound-electron $g$-factors are from Ref.~\cite{Tiesinga:2021myr} (see references therein). The predictions for the \hdplus and \hepbar  transition frequencies are from Ref.~\cite{Korobov:2017tvt,Korobov:2021} and~\cite{Korobov:2008jhd,Korobov:2013mvu,Korobov:2014voa}, respectively, and are updated with the latest CODATA recommended values, see Sec.~\ref{sec:DATA22dataset} for details. 

The NP contributions to atomic and molecular ion transition frequencies are obtained using (time-independent) first-order perturbation theory~\cite{Cohen-Tannoudji:101367,messiah_2014}.
We use exact nonrelativistic wavefunctions for hydrogen-like atoms 
and very precise nonrelativistic numerical ones from a variational method of Ref.~\cite{Korobov:2000} for \hdplus and $\bar p$He. 
Expectation values of the Yukawa potentials in Eq.~\eqref{eq:VNP} 
are calculated for a grid of $\mNP$ values, taking advantage of the fact that their matrix elements in the chosen basis can be obtained in an analytical form. 
The precision is limited to $\cO(\alphaEM^2)$ because of the neglected relativistic corrections to the wavefunction.
The NP contribution to the free-electron $(g-2)_e$ arises at one-loop~\cite{Jackiw:1972jz,Jegerlehner:2009ry}, 
 while for bound electrons we include an additional tree-level contribution from electron-nucleus interaction~\cite{Debierre:2020}.
Finally, we assume NP to have negligible effects in atom recoil measurements as well as relative atomic mass measurements from cyclotron frequency measurements in Penning traps.

We pay particular attention to the possible degeneracy between the determination of SM and NP parameters. 
In the $\mNP\to0$ limit, the dark photon is completely degenerate with the QED photon, since couplings of the two are aligned, $q_i=Q_i$, and thus only the combination $\alpha+\alpha_\phi$ can be determined from data.
This degeneracy should be retained in the theoretical predictions \eqref{eq:th:predict}, which  
in principle requires calculating NP effects to the same very high order as the SM. 
We propose an alternative procedure, which uses the state-of-the-art SM calculations but requires NP contribution only at LO in $\alphaNP$, and reproduces the correct $q_i\to Q_i$,  $\mNP a_0 \ll 1$ limit, where $a_0\equiv \alpha/(4\pi R_\infty)=(\alpha m_e)^{-1}$ is the Bohr radius. 

For {\em light vectors}  we rewrite the NP potential in Eq.~\eqref{eq:VNP} as the sum of the Coulomb-like potential with QED coupling $Q_i$ plus the remainder, 
\begin{align}
    \label{eq:Vtilde}
    V^{ij}_\NP(r)
=   \alphaNP  \frac{Q_i Q_j}{r}
    +\widetilde{V}_\NP^{ij}(r) \,,
\end{align}
where $\widetilde{V}^{ij}_\NP(r)\equiv \alphaNP \left(q_i q_j e^{-\mNP r}-Q_iQ_j \right)/r$.
The theory predictions are evaluated at LO in $\widetilde{V}_\NP(r)$, while the NP Coulomb term and the related relativistic corrections are evaluated to the same order as the SM, which amounts to replacing $\alphaEM \to \alphaEM + \alpha_{\phi}$  in the SM predictions.  
 For any observable $\cO$ the theoretical prediction is then
\begin{align}
    \label{eq:Ovector}
    \cO
= \cO_{\SM}\left( \alpha+ \alphaNP\right) 
    \!+\!
    \widetilde{\cO}_\NP\left(\alpha+\alphaNP,\alphaNP,\mNP\right)  \, , 
\end{align}
where $\cO_{\SM}$ is the SM contribution now expressed as a function of $\alpha+\alphaNP$ and $\widetilde{\cO}_\NP$ is the NP contribution from $\widetilde{V}_\NP$. 
 In the $m_\phi\to 0, q_i\to Q_i$ limit the potential $\widetilde{V}_\NP$ vanishes, 
 and all theory predictions are the SM ones, but shifted by $\alpha\to \alpha +\alpha_\phi$. For massive dark photon with $\mNP a_0\ll 1$ the leading effect of $\widetilde{V}_\NP$ is parametrically $\widetilde{\cO}_\NP\propto \mNP^2$. 
 Note that for massless $B-L$ the potential $\tilde V_{\rm NP}$ vanishes in hydrogen but not in deuterium where $\widetilde{\cO}_\NP\propto q_n$, thus breaking the degeneracy between the SM and NP contributions when $\mNP\to0$.

For {\em light scalars} there is no degeneracy with QED in the massless mediator limit; it is lifted by relativistic corrections. 
We can use directly the state-of-the-art SM predictions, and simply add to them the NP contribution due to the potential ~\eqref{eq:VNP} at LO, without any special treatments, see also Sec.~\ref{sec:masslessNP}.

%%%%%%%%%%%%%%%%%%%%%%%%%%%%%%%%%%%%%%%%%%%%%%%%%%%%%%%%%%%%%%%%%%%%%%
\section{Results}
\label{sec:result}
%%%%%%%%%%%%%%%%%%%%%%%%%%%%%%%%%%%%%%%%%%%%%%%%%%%%%%%%%%%%%%%%%%%%%%

\begin{figure}[t]
    \centering
    \includegraphics[width=0.8\columnwidth]{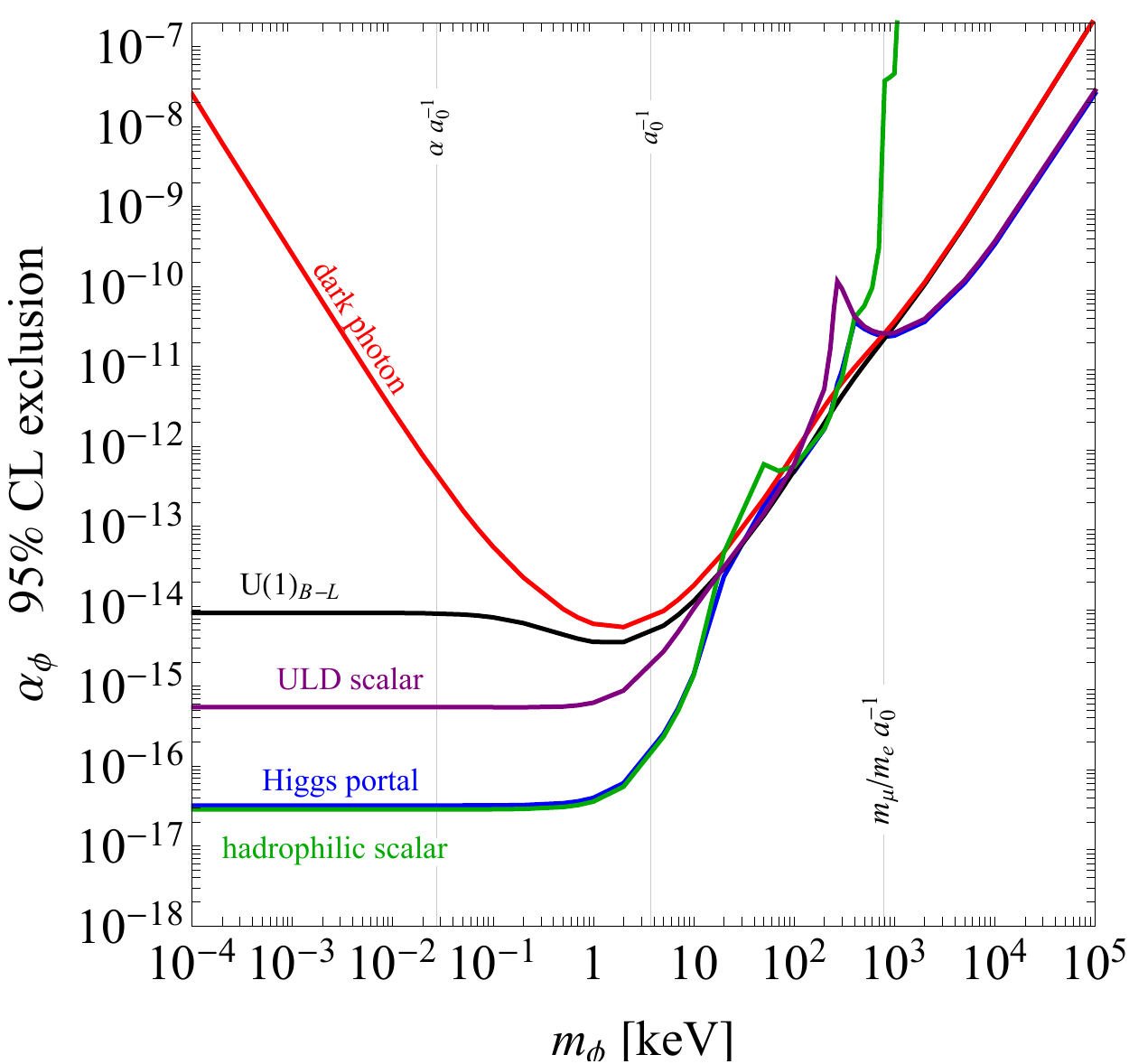}
    \caption{
   The  $95\%$ CL bounds on the NP coupling constant $\alphaNP$ as a function of the new boson's mass $\mNP$ for the benchmark NP models of Sec.~\ref{sec:NPModels}, as indicated. Other model-dependent constraints may apply (see text).  
}
    \label{fig:NPBound}
\end{figure}

First, we perform the control fit, {\it i.e.}, the least-squares adjustment assuming SM, based on the CODATA18 dataset with inflated experimental uncertainties when there are tensions in the data~\cite{Tiesinga:2021myr}, see also Sec.~\ref{sec:CODATA2018dataset}.
Resulting $\chi^2$ per degree-of-freedom~(dof) is $\chi^2_{\rm SM}/\nu_{\rm dof}\simeq 0.95$ ($\nu_{\rm dof}=78-44=34$), indicating an overall good description by the SM, and the use of correct expansion factors. 
The output $g_{\rm SM}$ values and relative uncertainties (see Table~\ref{tab:CODATA18adjus})
are in excellent agreement ($\lesssim0.2\,\sigma$) with the latest CODATA recommended values~\cite{Tiesinga:2021myr}, 
validating our procedure. 

Next, we perform adjustments based on the DATA22 dataset, assuming either the SM or one of the NP benchmark models in Sec.~\ref{sec:NPModels}. 
We do not inflate experimental errors,
since mild tensions in the data could be a hint of NP. 
The SM-only hypothesis still describes the data relatively well, with $\chi^2_{\rm SM}/\nu_{\rm dof}\simeq 1.4$ ($\nu_{\rm dof}=102-62=40$), despite known tensions in the proton charge radius puzzle data and the recent hydrogen $2S_{1/2}-8D_{5/2}$ transition~\cite{Brandt:2021yor}.

Figure~\ref{fig:NPBound} shows the 95\,\%~confidence level~(CL) upper bounds on $\alphaNP$ as function of $m_\phi$ for the NP benchmark models, Sec.~\ref{sec:NPModels}. 
The strongest exclusion is always reached around $m_\phi\sim a_0^{-1}\sim 4\,\keV$, and stays roughly constant for lighter $m_\phi$ (except for dark photon due to degeneracy with QED in the $m_\phi\to0$ limit, see Sec.~\ref{sec:fit}).
Deuterium observables translate to a $\sim2\times$ stronger bound on $B-L$ at $m_\phi\sim a_0^{-1}$, compared to dark photon.  
 The significantly stronger bounds on the Higgs portal and hadrophilic scalar for $\mNP \lesssim 10\,\keV$ are due to the $\sim \kappa_p m_p/m_e\simeq 500$ enhancement in inter-nucleon interactions (compared to electron--nucleon potential), affecting the \hdplus~observables.  
For heavier NP, $\mNP a_0\gtrsim 1 \; (m_\mu/m_e)$ in hydrogen (muonic hydrogen), the interaction is point-like, with suppressed electron (muon) wave function overlap, and the bounds decouple as $\propto 1/\mNP^2$ (and more quickly for hadrophilic scalar). 
The bounds are stronger for Higgs portal and ULD scalar due to $\sim m_\mu/m_e\simeq 200$ enhanced effects in muonic hydrogen.

\begin{figure}[t]
    \centering  
    \includegraphics[width=0.85\columnwidth]{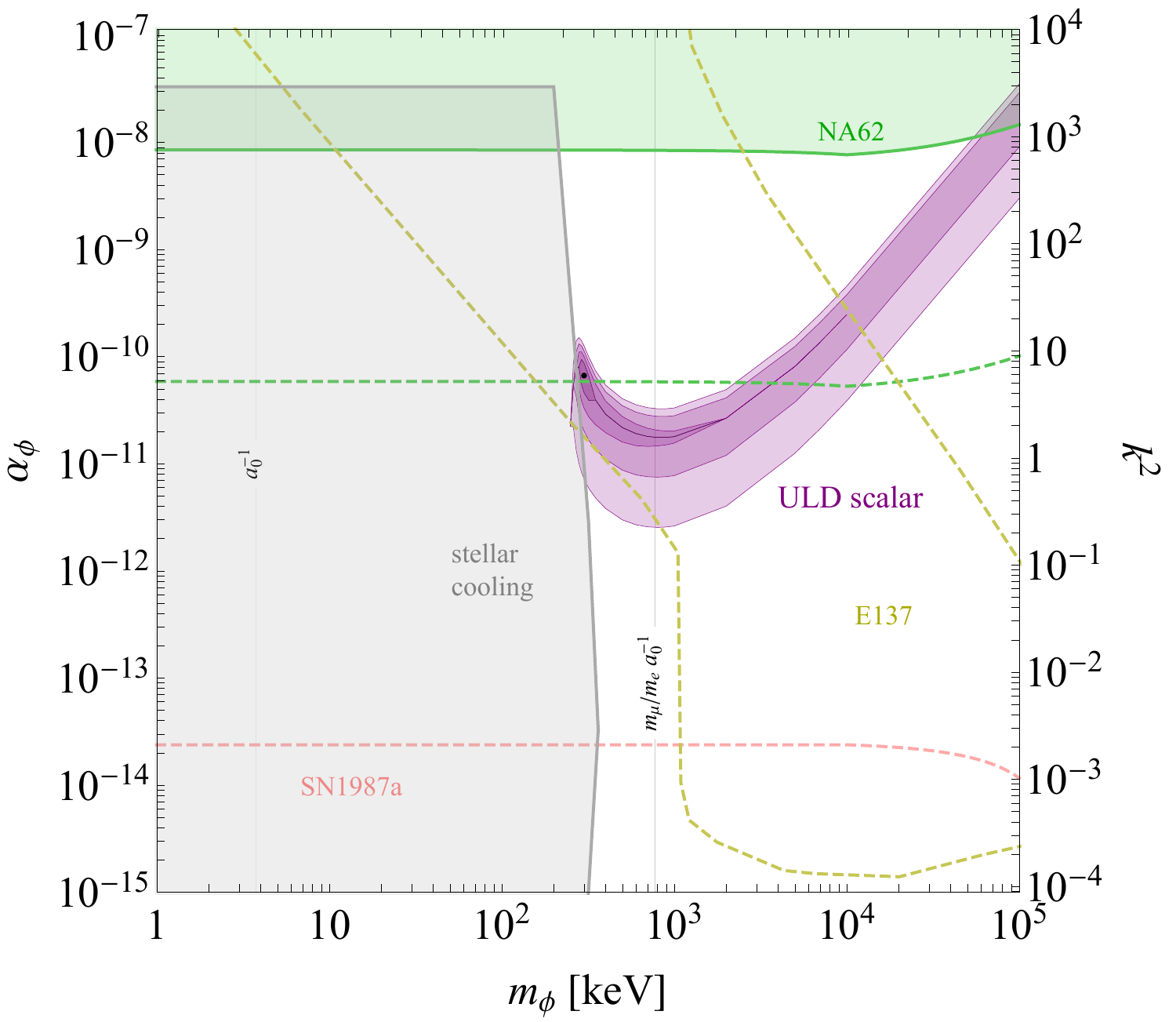}
    \caption{The constraints on ULD scalar in the $\alphaNP, \mNP$ plane, with purple-shaded $1,\,2,\,3,\,4\,\sigma$ CL regions favored by the DATA22 dataset (black dot is the best-fit point).
    Exclusions are by SN1987a~\cite{Raffelt:2012sp,Batell:2018fqo} (below the pink line, absent if $\phi$ invisible decay dominates), NA62 $K^+\to\pi^+X_{\rm inv}$ search~\cite{NA62:2020xlg} (green, the dashed line is a naive NNLO estimate), stellar cooling~\cite{Hardy:2016kme} (gray), and  E137~\cite{Bjorken:1988as,Liu:2016mqv} (between yellow dashed lines, absent if $\phi$ invisible decay dominates).
    }
    \label{fig:alphaNP-mNP}
\end{figure}

For $m_\phi a_0\gtrsim m_\mu/m_e$ the Higgs portal and ULD scalar are statistically preferred over the SM at the $\sim4\,\sigma$ and $\sim5\,\sigma$ level, respectively. 
Figure~\ref{fig:alphaNP-mNP} shows the preferred region for the ULD scalar, around the best-fit point, $\mNP=300\,\keV$ and $\alphaNP=6.7\times 10^{-11}$. 
This NP hint  
is supported mostly from the recent measurements of the hydrogen $2S_{1/2}-8D_{5/2}$ and $1S_{1/2}-3S_{1/2}$ transitions~\cite{Brandt:2021yor,Grinin:2020}, as well as muonic deuterium, {\it cf.} Sec. \ref{sec:NP:vs:SM}. 
While these tensions between data and the SM prediction are not new, 
our analysis shows that all tensions can be significantly ameliorated when including NP interactions due to a single light scalar. The favored NP mass is close to the (inverse) Bohr radius of muonic atoms, $a_0^{-1}\times m_\mu/m_e\sim\MeV$, due to the large muon-electron coupling ratio in these models, contrasting with scalars having weaker or vanishing coupling to muons (see Sec.~\ref{sec:further:NP} and~\cite{Brandt:2021yor}). 
However, other constraints require the scalar to have rather a nontrivial pattern of couplings, see Sec~\ref{sec:app-ULD}. 
For ULD scalar the E137~\cite{Bjorken:1988as,Liu:2016mqv} bounds are evaded since $\phi$ decays predominantly to an invisible dark sector. 
Since $\phi$ couples to up quarks and not directly to heavy quarks and gluons, the bound from NA62 search for $K^+\to\pi^+\phi$~\cite{NA62:2020xlg} is weakened~\cite{Batell:2018fqo,WorkInProgress}. 
Finally, the minimal ULD model induces a too large contribution to $(g-2)_\mu$, however this can be suppressed  
in less minimal versions with a custodial symmetry~\cite{Balkin:2021rvh}. 
\begin{figure}[tb]
     \centering
     \includegraphics[width=0.8\columnwidth]{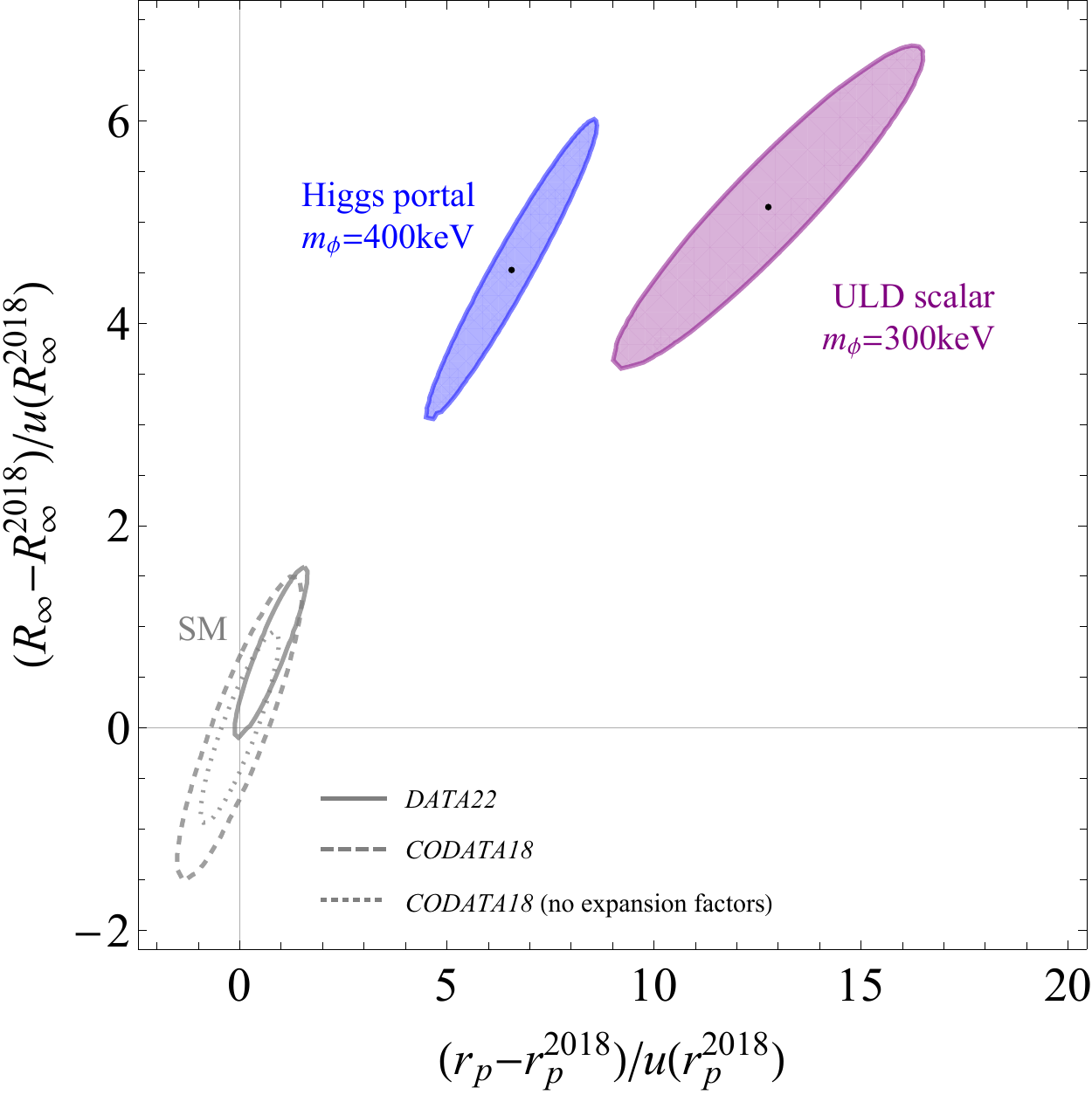}
     \caption{The 68$\%$~CL regions for simultaneous determinations of the
 Rydberg constant $R_\infty$ and the proton radius $r_p$ assuming either the SM-only hypothesis (gray) or including putative NP contributions from a  $400\,$keV  Higgs portal scalar (blue) or 300\,keV ULD scalar (purple). The solid lines use DATA22 dataset, the dashed (dotted) lines the CODATA18 dataset with (without) errors inflated by expansion factors. Both $R_\infty$ and $r_p$ are shown in terms of normalized deviations from the central values of the CODATA 2018 analysis, Ref.~\cite{Tiesinga:2021myr}.
 }
     \label{fig:Correlation}
 \end{figure} 

The presence of NP also impacts the determination of the fundamental constants in the SM.
Figure~\ref{fig:Correlation} shows the $68\,\%$~CL determination of $r_p$ and $R_\infty$, subtracting the CODATA 2018 recommended values and normalizing to respective errors.  The  SM-parameter uncertainties increase in the presence of NP and the central values shift outside the nominal SM ellipse, shown explicitly in Fig.~\ref{fig:Correlation} for the Higgs portal and ULD scalar model. Because of the degeneracy with the photon the uncertainty on $\alphaEM$ in the dark photon model increases as $1/m_\phi^2$ for masses below $10\,$eV (see Fig.~\ref{fig:uncertainties}) and eventually becomes comparable to $\alphaEM$ itself for $m_\phi\sim 0.1\,$meV, while $\alphaEM+\alphaNP$ remains well constrained.\\
%

%%%%%%%%%%%%%%%%%%%%%%%%%%%%%%%%%%%%%%%%%%%%%%%%%%%%%%%%%%%%%%%%%%%%%
\section{Conclusions}
\label{sec:conclusions}
%%%%%%%%%%%%%%%%%%%%%%%%%%%%%%%%%%%%%%%%%%%%%%%%%%%%%%%%%%%%%%%%%%%%%
Extracting bounds on light NP from a global fit to spectroscopic and other precision data requires both SM and NP parameters to be determined simultaneously. 
The possibility of NP contributions changes the extracted allowed ranges of SM parameters, a change that can be quite substantial, see Fig. \ref{fig:Correlation}. 
Furthermore, we provided a prescription to consistently include NP corrections from light vectors. It requires calculations of NP contribution only at leading order, and recovers the expected degeneracy between dark photon and QED in the massless mediator limit. 

At present, spectroscopic data show tensions that could either be due to unknown or under-appreciated systematics, or to light NP.  
We showed that the $\sim 4\sigma$ anomaly in data can be explained by a flavor non-universal light scalar model.

%%%%%%%%%%%%%%%%%%%%%%%%%%%%%%%%%%%%%%%%
\begin{acknowledgments}
We would like to thank Dmitry Budker and Gilad Perez for useful discussions and
comments on the manuscript.
The work of CD is supported by the CNRS IRP NewSpec. 
The work of TK is supported by the Japan Society for the Promotion of Science (JSPS) Grant-in-Aid for Early-Career Scientists (Grant No.\,19K14706) and the JSPS Core-to-Core Program (Grant No.\,JPJSCCA20200002). 
The work of YS is supported by grants from the NSF-BSF (No. 2018683), the ISF (No. 482/20), the BSF (No. 2020300) and by the Azrieli foundation.
JZ acknowledges support in part by the DOE grant de-sc0011784. 
This work was also supported in part by the European Union’s Horizon 2020 research and innovation programme, project STRONG2020, under grant agreement No. 824093.
\end{acknowledgments}
%%%%%%%%%%%%%%%%%%%%%%%%%%%%%%%%%%%%%%%%%%%%%%%%%%%%%%%%%%%%%%%%%%%%%%

%%%%%%%%%%%%%%%%%%%%%%%%%%%%%%%%%%%%%%%%%%%%%%%%%%%%%%%%%%%%%%%%%%%%%%
%\bibliographystyle{utphys}
\bibliographystyle{utphys28mod}
\bibliography{CODATA_NP_bib}
%%%%%%%%%%%%%%%%%%%%%%%%%%%%%%%%%%%%%%%%%%%%%%%%%%%%%%%%%%%%%%%%%%%%%%

\input{supplement}

\end{document}

%% file: supplement.tex
\clearpage
\onecolumngrid
%\appendix 

\setcounter{equation}{0}
\setcounter{figure}{0}
\setcounter{table}{0}
\setcounter{section}{0}
\makeatletter
\renewcommand{\theequation}{S\arabic{equation}}
\renewcommand{\thefigure}{S\arabic{figure}}
\renewcommand{\thetable}{S\arabic{table}}
\renewcommand{\thesection}{S\arabic{section}}
%%%%%%%%%%%%%%%%%%%%%%%%%%%%%%%%%%%%%%%%%%%%%%%%%%%%%%%%%%%%%%%%%%%%%%
\begin{center}
 \large{\bf 
 Self-consistent extraction of spectroscopic bounds on light new physics
 }\\
 Supplemental Material
\end{center}
%%%%%%%%%%%%%%%%%%%%%%%%%%%%%%%%%%%%%%%%%%%%%%%%%%%%%%%%%%%%%%%%%%%%%%
\begin{center}
C\'edric Delaunay, 
Jean-Philippe Karr,
Teppei Kitahara,
Jeroen C. J. Koelemeij,
Yotam Soreq, and 
Jure Zupan
\end{center}

In the supplemental material we give further details on 
the linearized least squares method (Sec.~\ref{sec:app:fit}), the CODATA18 dataset (Sec.~\ref{sec:CODATA2018dataset}), the DATA22 dataset (Sec.~\ref{sec:DATA22dataset}), the new physics benchmarks models (Sec.~\ref{sec:further:NP}), new physics improvements relative to the SM (Sec.~\ref{sec:NP:vs:SM}), the extraction of fundamental constants in the presence of NP (Sec.~\ref{sec:app:extraction:NP}), and the $m_\phi\to0$ limit (Sec.~\ref{sec:masslessNP}).

\section{The linearized least squares method}
\label{sec:app:fit}
We use the method of linearized least squares employed by the CODATA to determine both the fundamental and NP constants. 
For completeness we summarize below briefly the fitting procedure, while further details can be found in Appendix~E of Ref.~\cite{Mohr:2000ie}.

The $N$ input data $w_i$ are either measured quantities or estimated errors on theoretical calculations (for these the central values are taken to be zero), and are listed in  Tables \ref{tab:inputsA}-\ref{tab:inputsD} and \ref{tab:inputsNEW}-\ref{tab:HDuncertNEW}, \ref{tab:3buncert}, see also Sections \ref{sec:CODATA2018dataset} and \ref{sec:DATA22dataset} for a detailed discussion of the two datasets we use.  
The predictions for $w_i$ are functions of $M$ fitted parameters (adjusted constants) $z_j$, 
\begin{align}
    \label{eq:obseq}
    w_i \dot{=} f_i(z)\equiv f_i(z_1,z_2,\dots,z_M)\,.   
\end{align}
The dotted equality sign means that the left and right hand sides are not equal in general, since the set of equations is usually overdetermined ($N\geq M$), but should agree within experimental errors, if the assumed physics model and measurements are correct. 
Most of the $f_i$ are nonlinear functions, which we linearize, {\it i.e.}, Taylor expand around some initial values $s_j$, chosen close enough to the expected values of $z_j$
such that the higher-order terms can be neglected, 
\begin{align}
    w_i 
    \dot{=} 
    f_i(s) + \sum_{j=1}^{M}\frac{\partial f_i(s)}{\partial z_j}(z_j-s_j)+\cdots\,.
\end{align}

Linearization allows for more efficient determination of adjusted constants. 
Observational equations \eqref{eq:obseq} are now a set of overdetermined linear equations. In matrix notation these are
\begin{align}
    Y\dot{=} AX\,,    
\end{align}
where $Y$ is an $N$-dimensional vector of $y_i\equiv w_i-f_i(s)$, $A$ is an $N\times M$ matrix with elements $a_{ij}\equiv \partial f_i(s)/\partial z_j$, while $X$ is an $M$-dimensional vector of $x_j\equiv z_j-s_j$. The best fit value of $X$ is obtained by minimizing the $\chi^2$ function,
\begin{align}
    \chi^2= (Y-AX)^T V^{-1}(Y-AX)\,,
\end{align}
where $V$ is the $N\times N$ covariance matrix constructed from variances $u_{ii}=u_i^2$ and covariances $u_{ij}\equiv u(w_i,w_j)$ for inputs $w_{i,j}$ (here $u_i\equiv u(w_i)$ is the standard uncertainty). Updated correlation coefficients for the DATA22 dataset are listed in Tables \ref{tab:HDcorrelNEW}, \ref{tab:3bcorrel}.
The result of the fit, vector $\hat{X}$ with components $\hat{x}_j$ that minimizes $\chi^2$, is given by
%  is 
% 
\begin{align}
    \hat{X} = G A^TV^{-1}Y\,,    
\end{align}
where $G\equiv (A^T V^{-1} A)^{-1}$ is the covariance matrix of $\hat X$, {\it i.e.}, an $M\times M$ matrix that captures the uncertainties and correlations of the adjusted constants $z_j$.

The solution $\hat{X}$ is only an approximation of the exact solution to the nonlinear problem. 
More precise values of the adjusted fundamental constants are obtained by iterating the above procedure using $s_j+\hat{x}_j$ as starting values, until the new $\hat{x}_j$'s are negligibly small compared to their estimated uncertainties. 
Similarly to the CODATA, we use the convergence condition 
\begin{align}
    \sum_{j=1}^{M}\frac{\hat {x}_j^2}{u^2(\hat x_j)}< 10^{-20}\,.   
\end{align}
In practice, this condition is satisfied after a few iterations. 

Finally, we define for each input datum  a normalized residual 
\begin{align}
    \label{eq:Ridef}
    R_i\equiv  \frac{w_i-f_i(\hat z)}{u_i}\,, 
\end{align}
where $\hat z$ denotes the final value of the adjusted constants.  Ignoring correlations, the contribution of each input datum $i$ to the minimal $\chi^2$ is $R_i^2$.

\section{The CODATA18 dataset}
\label{sec:CODATA2018dataset}

%%%%%%%%%%%%%%%%%%%%%%%%%%%%%%%%%%%%%%%%%%%%%%%%%%%%%%%%%%%%%%%%%%%%%%
\begin{table}[t]
    \centering
    \renewcommand{\arraystretch}{1.2}
    \begin{tabular}{cccc}
    \toprule\hline
Label & \multicolumn{1}{c}{Input datum} & \multicolumn{1}{c}{Value (kHz)} & \multicolumn{1}{c}{Rel. uncert.} \\
\midrule
A1 & $\nuh(2S_{1/2}-4S_{1/2})-\frac{1}{4}\nuh(1S_{1/2}-2S_{1/2})$ & \num{4 797 338(10)} & \num{2.1e-6}  \\
A2 & $\nuh(2S_{1/2}-4D_{5/2})-\frac{1}{4}\nuh(1S_{1/2}-2S_{1/2})$ & \num{6 490 144(24)} & \num{3.7e-6}  \\
A3 & $\nud(2S_{1/2}-4S_{1/2})-\frac{1}{4}\nud(1S_{1/2}-2S_{1/2})$ & \num{4 801 693(20)} & \num{4.2e-6}  \\
A4 & $\nud(2S_{1/2}-4D_{5/2})-\frac{1}{4}\nud(1S_{1/2}-2S_{1/2})$ & \num{6 494 841(41)} & \num{6.3e-6}  \\
A5 & $\nud(1S_{1/2}-2S_{1/2})-\nuh(1S_{1/2}-2S_{1/2})$ & \num{670 994 334.606(15)} & \num{2.2e-11}  \\
A6 & $\nuh(1S_{1/2}-2S_{1/2})$   & \num{2 466 061 413 187.035(10)} & \num{4.2e-15}  \\
A7 & $\nuh(1S_{1/2}-2S_{1/2})$   & \num{2 466 061 413 187.018(11)} & \num{4.4e-15}  \\
A8 & $\nuh(1S_{1/2}-3S_{1/2})$   & \num{2 922 743 278 659(17)} & \num{5.8e-12}  \\
A9 & $\nuh(2S_{1/2}-4P)$         & \num{616520931626.8}(2.3) & \num{3.7e-12}  \\
A10 & $\nuh(2S_{1/2}-8S_{1/2})$  & \num{770 649 350 012.0}(8.6) & \num{1.1e-11}  \\
A11 & $\nuh(2S_{1/2}-8D_{3/2})$  & \num{770 649 504 450.0}(8.3) & \num{1.1e-11}  \\
A12 & $\nuh(2S_{1/2}-8D_{5/2})$  & \num{770 649 561 584.2}(6.4) & \num{8.3e-12}  \\
A13 & $\nud(2S_{1/2}-8S_{1/2})$  & \num{770 859 041 245.7}(6.9) & \num{8.9e-12}  \\
A14 & $\nud(2S_{1/2}-8D_{3/2})$  & \num{770 859 195 701.8}(6.3) & \num{8.2e-12}  \\
A15 & $\nud(2S_{1/2}-8D_{5/2})$  & \num{770 859 252 849.5}(5.9) & \num{7.7e-12}  \\
A16 & $\nuh(2S_{1/2}-12D_{3/2})$ & \num{799 191 710 472.7}(9.4) & \num{1.2e-11}  \\
A17 & $\nuh(2S_{1/2}-12D_{5/2})$ & \num{799 191 727 403.7}(7.0) & \num{8.7e-12}  \\
A18 & $\nud(2S_{1/2}-12D_{3/2})$ & \num{799 409 168 038.0}(8.6) & \num{1.1e-11}  \\
A19 & $\nud(2S_{1/2}-12D_{5/2})$ & \num{799 409 184 966.8}(6.8) & \num{8.5e-12}  \\
A20 & $\nuh(2S_{1/2}-6S_{1/2})-\frac{1}{4}\nuh(1S_{1/2}-3S_{1/2})$ & \num{4 197 604(21)} & \num{4.9e-6}  \\
A21 & $\nuh(2S_{1/2}-6D_{5/2})-\frac{1}{4}\nuh(1S_{1/2}-4S_{1/2})$ & \num{4 699 099(10)} & \num{2.2e-6}  \\
A22 & $\nuh(1S_{1/2}-3S_{1/2})$  & \num{2 922 743 278 678(13)} & \num{4.4e-12}  \\
A23 & $\nuh(1S_{1/2}-3S_{1/2})$  & \num{2 922 743 278 671.5}(2.6) & \num{8.9e-13}  \\
A24 & $\nuh(2S_{1/2}-4P_{1/2})-\frac{1}{4}\nuh(1S_{1/2}-2S_{1/2})$ & \num{4 664 269(15)} & \num{3.2e-6}  \\
A25 & $\nuh(2S_{1/2}-4P_{3/2})-\frac{1}{4}\nuh(1S_{1/2}-2S_{1/2})$ & \num{6 035 373(10)} & \num{1.7e-6}  \\
A26 & $\nuh(2S_{1/2}-2P_{3/2})$  & \num{9 911 200(12)} & \num{1.2e-6}  \\
A27 & $\nuh(2P_{1/2}-2S_{1/2})$  & \num{1 057 862(20)} & \num{1.9e-5}  \\
A28 & $\nuh(2P_{1/2}-2S_{1/2})$  & \num{1 057 845.0}(9.0) & \num{8.5e-6}  \\
A29 & $\nuh(2P_{1/2}-2S_{1/2})$  & \num{1 057 829.8}(3.2) & \num{3.0e-6}  \\
\hline\bottomrule
    \end{tabular}
    \caption{Input data of the electronic hydrogen and deuterium measurements for the CODATA18 dataset, taken from Table X of Ref.~\cite{Tiesinga:2021myr}. }
    \label{tab:inputsA}
\end{table}
%%%%%%%%%%%%%%%%%%%%%%%%%%%%%%%%%%%%%%%%%%%%%%%%%%%%%%%%%%%%%%%%%%%%%%

%%%%%%%%%%%%%%%%%%%%%%%%%%%%%%%%%%%%%%%%%%%%%%%%%%%%%%%%%%%%%%%%%%%%%%
\begin{table}[h]
    \centering
    \renewcommand{\arraystretch}{1.2}
    \begin{tabular}{cccc}
    \toprule\hline
Label & \multicolumn{1}{c}{Input datum} & \multicolumn{1}{c}{Value (kHz)} & \multicolumn{1}{c}{Rel. uncert.} \\
\midrule
B1  & $\dH(1S_{1/2})/h$ & \num{0.0}(\num{1.6}) & \num{4.9e-13} \\
B2  & $\dH(2S_{1/2})/h$ & \num{0.00(20)}   & \num{2.4e-13} \\
B3  & $\dH(3S_{1/2})/h$ & \num{0.000(59)}  & \num{1.6e-13} \\
B4  & $\dH(4S_{1/2})/h$ & \num{0.000(25)}  & \num{1.2e-13} \\
B5  & $\dH(6S_{1/2})/h$ & \num{0.000(12)}  & \num{1.3e-13} \\
B6  & $\dH(8S_{1/2})/h$ & \num{0.0000(51)} & \num{9.9e-14} \\

B7  & $\dH(2P_{1/2})/h$ & \num{0.0000(39)} & \num{4.8e-15} \\
B8  & $\dH(4P_{1/2})/h$ & \num{0.0000(16)} & \num{7.6e-15} \\
B9  & $\dH(2P_{3/2})/h$ & \num{0.0000(39)} & \num{4.8e-15} \\
B10  & $\dH(4P_{3/2})/h$ & \num{0.0000(16)} & \num{7.6e-15} \\
B11  & $\dH(8D_{3/2})/h$ & \num{0.000\,000(13)} & \num{2.6e-16} \\
B12  & $\dH(12D_{3/2})/h$ & $0.000\,0000(40)$ & \num{1.8e-16} \\
B13  & $\dH(4D_{5/2})/h$ & \num{0.000\,00(17)} & \num{8.2e-16} \\
B14  & $\dH(6D_{5/2})/h$ & \num{0.000\,000(58)} & \num{6.3e-16} \\
B15  & $\dH(8D_{5/2})/h$ & \num{0.000\,000(22)} & \num{4.2e-16} \\
B16  & $\dH(12D_{5/2})/h$ & $0.000\,0000(64)$ & \num{2.8e-16} \\
B17 & $\dD(1S_{1/2})/h$ & \num{0.0}(\num{1.5})   & \num{4.5e-13} \\
B18 & $\dD(2S_{1/2})/h$ & \num{0.00(18)}   & \num{2.2e-13} \\
B19 & $\dD(4S_{1/2})/h$ & \num{0.000(23)}  & \num{1.1e-13} \\
B20 & $\dD(8S_{1/2})/h$ & \num{0.0000(49)} & \num{9.6e-14} \\
B21 & $\dD(8D_{3/2})/h$ & $0.000\,0000(95)$ & \num{1.8e-16} \\
B22 & $\dD(12D_{3/2})/h$ & $0.000\,0000(28)$ & \num{1.2e-16} \\
B23 & $\dD(4D_{5/2})/h$ & \num{0.000\,00(15)} & \num{7.5e-16} \\
B24 & $\dD(8D_{5/2})/h$ & \num{0.000\,000(19)} & \num{3.8e-16} \\
B25 & $\dD(12D_{3/2})/h$ & $0.000\,0000(58)$ & \num{2.5e-16} \\
\hline\bottomrule
    \end{tabular}
    \caption{Input data for the additive energy corrections accounting for missing contributions to the theoretical description of H and D energy levels for the CODATA18 dataset, taken from Table~VIII of Ref.~\cite{Tiesinga:2021myr}. Relative uncertainties are with respect to the binding energy.}
    \label{tab:HDuncert}
\end{table}

%%%%%%%%%%%%%%%%%%%%%%%%%%%%%%%%%%%%%%%%%%%%%%%%%%%%%%%%%%%%%%%%%%%%%%
\begin{table}[tbh]
    \centering
    \renewcommand{\arraystretch}{1.2}
    \begin{tabular}{cccc}
    \toprule\hline
Label & \multicolumn{1}{c}{Input datum} & \multicolumn{1}{c}{Value} & \multicolumn{1}{c}{Rel. uncert.} \\
\midrule
C1 & $ E_{\rm LS}(\mu \text{H})$ & \num{202.3706(23)} meV & \num{1.1e-5}  \\
C2 & $ E_{\rm LS}(\mu \text{D})$ & \num{202.8785(34)} meV & \num{1.7e-5}  \\
C7 & $ \delta E_{\rm LS}(\mu \text{H})$ & \num{0.0000(129)} meV & \num{6.4e-5}  \\
C8 &  $\delta E_{\rm LS}(\mu \text{D})$ & \num{0.0000(210)} meV & \num{1.0e-4}  \\
C9 & $r_p$      & \num{0.880(20)} fm & \num{2.3e-2}  \\
C10 & $r_d$ & \num{2.111(19)} fm & \num{9.0e-3}  \\
\hline\bottomrule
    \end{tabular}
    \caption{Input data of the muonic hydrogen and deuterium Lamb shifts (LS), as well as the proton and deuteron charge radii from the electron-proton and electron-deuteron scatterings
     for the CODATA18 dataset, taken from Table~XVIII of Ref.~\cite{Tiesinga:2021myr}. For the additive corrections C7 and C8, the relative uncertainty is with respect to the value of the corresponding theoretical quantity.
     }
    \label{tab:inputsC}
\end{table}
%%%%%%%%%%%%%%%%%%%%%%%%%%%%%%%%%%%%%%%%%%%%%%%%%%%%%%%%%%%%%%%%%%%%%%

%%%%%%%%%%%%%%%%%%%%%%%%%%%%%%%%%%%%%%%%%%%%%%%%%%%%%%%%%%%%%%%%%%%%%%
\begin{table}[tbh]
    \centering
    \renewcommand{\arraystretch}{1.2}
    \begin{tabular}{cccc}
    \toprule\hline
Label & \multicolumn{1}{c}{Input datum} & \multicolumn{1}{c}{Value} & \multicolumn{1}{c}{Rel. uncert.} \\
\midrule
D1 & $ a_{e} \equiv \frac{1}{2}(g-2)_e$     & \num{1.159 652 180 73(28)} $\times 10^{-3}$& \num{2.4e-10}  \\
D2 & $\delta_e$ & \num{0.000(18)e-12} & \num{1.5e-11}\\
\OLD{D3} & \OLD{ $h/m_{\text{Rb}} ({}^{87}\text{Rb}) $ }& \OLD{\num{4.591 359 2729(57)} $\times 10^{-9} \,\text{m}^2\text{s}^{-1}$ }& \OLD{\num{1.2e-9} }\\
D4 &  $h/m_{\text{Cs}} ({}^{133}\text{Cs}) $ & \num{3.002 369 4721(12)} $\times 10^{-9} \,\text{m}^2\text{s}^{-1}$& \num{4.0e-10}  \\
\OLD{D5} & \OLD{$ A_{\rm r} ({}^{87}\text{Rb})$} & \OLD{\num{86.909 180 5312(65)}} & \OLD{\num{7.4e-11}}   \\
\OLD{D6} & \OLD{$A_{\rm r} ({}^{133}\text{Cs}) $} & \OLD{\num{132.905 451 9610(86)}} & \OLD{\num{6.5e-11}} \\
D7 & $ \omega_{\rm s}/\omega_{\rm c} ({}^{12}\text{C}^{5+})$ & \num{4376.210 500 87(12)} & \num{2.8e-11} \\
D8 & $\Delta E_{\rm B} ({}^{12}\text{C}^{5+})/hc $ & \num{43.563 233(25)} $\times 10^{7}\, \text{m}^{-1}$ & \num{5.8e-7} \\
D9 & $\delta_\text{C}$ & $0.0(2.5)$ $\times 10^{-11}$ & \num{1.3e-11} \\
D10 & $\omega_{\rm s}/\omega_{\rm c} ({}^{28}\text{Si}^{13+}) $& \num{3912.866 064 84(19)} & \num{4.8e-11} \\
\OLD{D11} & \OLD{$A_{\rm r} ({}^{28}\text{Si}) $} & \OLD{\num{27.976 926 534 99(52)}} & \OLD{\num{1.9e-11}} \\
D12 & $\Delta E_{\rm B}  ({}^{28}\text{Si}^{13+}) /hc $ & \num{420.6467(85)} $\times 10^{7}\, \text{m}^{-1}$& \num{2.0e-5}  \\
D13 & $\delta_\text{Si}$ & $0.0(1.7)$ $\times 10^{-9}$ & \num{8.3e-10} \\
\OLD{D14} & \OLD{$ \omega_{\rm c}(\text{d})/\omega_{\rm c}({}^{12}\text{C}^{6+})$}     & \OLD{\num{0.992 996 654 743(20)}} & \OLD{\num{2.0e-11}}  \\
\OLD{D15} & \OLD{$\omega_{\rm c}({}^{12}\text{C}^{6+})/\omega_{\rm c}(p) $} & \OLD{\num{0.503 776 367 662(17)}} & \OLD{\num{3.3e-11}}   \\
\OLD{D19} & \OLD{$A_{\rm r} ({}^{1}\text{H}) $} & \OLD{\num{1.007 825 032 241(94)}} & \OLD{\num{9.3e-11}}  \\
D21 & $ \Delta E_{\rm B}  ({}^{1}\text{H}^{+}) /hc$ & \num{1.096 787 717 4307(10)} $\times 10^{7}\, \text{m}^{-1}$ & \num{9.1e-13}   \\
\OLD{D23} & \OLD{$ \Delta E_{\rm B}  ({}^{12}\text{C}^{6+}) /hc$} & \OLD{\num{83.083 850(25)} $\times 10^{7}\, \text{m}^{-1}$} & \OLD{\num{3.0e-7}} \\
\hline \bottomrule
    \end{tabular}
    \caption{Input data relevant for fundamental constants other than the Rydberg constant and nuclear charge radii 
    for the CODATA18 dataset, taken from  Table~XXI of Ref.~\cite{Tiesinga:2021myr}. The D1-D13 inputs are relevant for the fine-structure constant and the electron mass, while the D14-D23 ones are relevant for the proton and deuteron masses. For the additive corrections D2, D9 and D13 of the $a_e$ and bound $g$-factors in carbon (C) and silicium (Si), respectively, the relative uncertainty is relative to the value of the corresponding theoretical quantity.}
    \label{tab:inputsD}
\end{table}
%%%%%%%%%%%%%%%%%%%%%%%%%%%%%%%%%%%%%%%%%%%%%%%%%%%%%%%%%%%%%%%%%%%%%%

\begin{table}[t]
  \renewcommand{\arraystretch}{1.2}
    \centering
    \begin{tabular}{cc cc c ccc }
    \hline\hline
\multicolumn{2}{c}{dataset}  &    \multicolumn{2}{c}{CODATA18} & &  \multicolumn{3}{c}{DATA22}\\
        constant & unit & value &  rel. uncert. &  & value &  rel. uncert. & shift \\ \hline
      $R_\infty c$   & Hz & $3.289\,841\,960\,2512(64)\times 10^{15}$ & $1.9\times 10^{-12}$ &  & $3.289\,841\,960\,2563(35)\times 10^{15}$ & $1.1\times 10^{-12}$ &  $+0.9\sigma$\\
      $r_p$ & fm &  $0.8414(19)$& $2.2\times 10^{-3}$ & &$0.8428(11)$ &$1.3\times 10^{-3}$ & $+0.8\sigma$\\
      $r_d$ & fm &  $2.128\,04(74)$ & $3.5\times 10^{-4}$& &$2.128\,59(43)$& $2.0\times 10^{-4}$ & $+0.8\sigma$\\
      $\alpha$ & & $7.297\,352\,569\,3(11)\times 10^{-3}$ & $1.5\times 10^{-10}$& &$7.297\,352\,564\,47(68)\times 10^{-3}$& $9.3\times10^{-11}$ & $-4.3\sigma$\\
      $A_{\rm r}(e)$ & u  & $5.485\,799\,090\,67(15)\times 10^{-4}$ & $2.8\times 10^{-11}$&&$5.485\,799\,090\,397(94)\times 10^{-4}$& $1.7\times10^{-11}$& $-1.7\sigma$\\
      $A_{\rm r}(p)$ & u & $1.007\,276\,466\,622(54) $&  $5.4\times 10^{-11}$&& $1.007\,276\,466\,596(14)$& $1.3\times 10^{-11}$& $-0.5\sigma$\\
      $A_{\rm r}(d)$ & u &$2.013\,553\,212\,744(41) $ &  $2.0\times 10^{-11}$&& $2.013\,553\,212\,542(15)$ & $7.4\times10^{-12}$ &    $-5.0\sigma$\\
      \hline\hline
    \end{tabular}
    \caption{The values and relative standard uncertainties of the main fundamental constants resulting from the least-squares adjustment based on the CODATA18 and DATA22 datasets without new physics. The symbol u denotes the unified atomic mass unit. The last column indicates the shift of the DATA22 value from the CODATA18 one in units of the CODATA18 uncertainty. 
    }
    \label{tab:CODATA18adjus}
\end{table}

The CODATA~2018 dataset contains all the inputs from Ref.~\cite{Tiesinga:2021myr} related to the determination of the Rydberg constant $\Rinf$, the proton and deuteron radii, $r_p$ and $r_d$, respectively, the fine-structure constant $\alphaEM$, and the relative atomic masses of the electron, proton, and deuteron: $A_{\rm r}(e)$, $A_{\rm r}(p)$ and $A_{\rm r}(d)$, respectively. 
The other observables and parameters included in the CODATA~2018 adjustment are very weakly correlated  with the selected data, and can be neglected for our purposes.  

The selected data include atomic transitions and non-spectroscopic observables, which depend on the main fundamental constants as follows. 
The inputs primarily used to determine $\Rinf$, $r_p$, and $r_d$ are: 
\begin{enumerate}
\item[A)]~measurements of transition frequencies in electronic hydrogen and deuterium, labeled as A$i$, $i=1,\ldots,29$, and listed in Table~\ref{tab:inputsA}, 
\item[B)]~the additive theory uncertainties on theory predictions for the relevant energy levels, labeled by B$i$, $i=1, \dots, 25$,  listed in Table~\ref{tab:HDuncert}, and 
\item[C)]~the inputs for muonic hydrogen and deuterium and electron scattering, labeled by C$i$, with $i=1,2,7,8$  and $i=9,10$, respectively (see Table~\ref{tab:inputsC}). 
\end{enumerate}
The dependence of an electronic transition frequency $\nu_{\SM}^i$ on physical constants is described, within the SM, by the simplified expression
\begin{align}
    \label{eq:nuSMi}
    \nu^i_{\SM} 
=   2c\Rinf \left[ 
    \frac{a_i}{1 + m_e/m_N}
    + b_i(\alphaEM) + c_i(r_N) \right] \,,
\end{align}
where $a_i \equiv 1/n^{\prime2}_i - 1/n_i^2$, with $n_i$, $n'_i$ the principal quantum numbers of the initial and the final state, respectively, while $m_N$ is the mass of the nucleus, and $i$ runs over all the transitions in Table~\ref{tab:inputsA}. 
The coefficient $b_i(\alphaEM)$ denotes higher-order relativistic and QED corrections, which take the form of a power series in $\alphaEM$ and $\ln(\alphaEM)$, with the leading term starting at order $\cO(\alphaEM^2)$. 
Finally, $c_i (r_N)$, where $r_N$ is the nuclear charge radius, denotes the finite-nuclear-size and nuclear-polarizability corrections, where the leading term is of order $(r_N/a_0)^2$. 
The Bohr radius $a_0$ is given by 
\begin{align}
    a_0^{-1}\equiv 4\pi \Rinf{}/\alphaEM \, .
\end{align}

Since the hydrogen transition frequencies depend only weakly on $m_e/m_N$ and $\alphaEM$, these parameters have to be determined by other means, using inputs listed in Table~\ref{tab:inputsD}, and labeled as D$i$, $i=1, \ldots, 23$.  
The fine-structure constant $\alphaEM$ is extracted from two different methods:  
(i)~a comparison of theoretical and experimental results for the anomalous magnetic moment of the electron~\cite{Hanneke:2008tm} and 
(ii)~atom-recoil experiments~\cite{Bouchendira:2011,Parker:2018vye} that measure the mass $m_A$ of atom $A$ in units of the Planck constant $h$, combined with values of the Rydberg constant and masses of the electron and atom $A$ in atomic mass units. 
Following Ref.~\cite{Tiesinga:2021myr}, the CODATA 2018 dataset therefore also includes the values of relative atomic masses for the relevant atoms. 
These are taken from the 2016 Atomic Mass Evaluation~(AME)~\cite{Huang:2017,Wang:2017} (see Table~\ref{tab:inputsD}). 

The electron relative atomic mass is obtained from the measurements of spin-flip 
and cyclotron 
frequencies in a hydrogenic ion $A$ using 
a theoretical calculation of the bound-electron $g$-factor $g_e(A)$. 
The atomic masses of the ions are deduced from the AME 2016 values of their neutral counterparts by subtracting the mass of the missing electrons and theoretically correcting for their binding energies~\cite{NIST_ASD}.

The relative atomic mass of proton is extracted from the AME 2016 value of the hydrogen atom (using the theoretical binding energy) along with the more recent measurement~\cite{Heisse:2017}; for the  relative atomic mass of deuteron, only the most recent measurement~\cite{Zafonte:2015} is taken into account. 
The correlation coefficients for the A, B and D datasets are listed in Table~IX and Table~XXII of Ref.~\cite{Tiesinga:2021myr}, respectively.

Finally, for the ease of comparison we follow Ref.~\cite{Tiesinga:2021myr}, and add expansion factors for the CODATA18 analysis, in the case where no new physics is considered. 
The expansion factors reduce the tension in the data: 
we multiply by a factor of $1.6$ the quoted errors for the proton radius data, \ie, to all the A, B, and C inputs, and a factor of $1.7$ for the proton mass data, \ie, the D15 
and D19 items.  
In the DATA22 analysis, on the other hand, we do not use the expansion factors.

Extracted values of $g_{\rm SM}$ using CODATA18 data and assuming no new physics contributions are listed in Table \ref{tab:CODATA18adjus}.

%%%%%%%%%%%%%%%%%%%%%%%%%%%%%%%%%%
\section{The DATA22 dataset}
\label{sec:DATA22dataset}
%%%%%%%%%%%%%%%%%%%%%%%%%%%%%%%%%%%%%%%

The DATA22 dataset contains the CODATA18 inputs, but with updated values for both  the experimental and theoretical inputs, Table \ref{tab:HDuncertNEW},
as well as the additional input data, Table \ref{tab:inputsNEW}. 
Together these then provide an improved sensitivity to NP effects.
 
To the list of hydrogen atom observables we added the two latest measurements of the $1S_{1/2}-3S_{1/2}$~\cite{Grinin:2020} and $2S_{1/2}-8D_{5/2}$~\cite{Brandt:2021yor} transition frequencies, and took into account recent theory improvements in the calculation of $nS_{1/2}$ energy levels~\cite{Yerokhin:2019,Karshenboim:2019a,Szafron:2019,Karshenboim:2019b}. 
The resulting slightly reduced theoretical uncertainties and their correlation coefficients are provided in Tables~\ref{tab:HDuncertNEW} and \ref{tab:HDcorrelNEW}, respectively. 

Among the data related to $\alphaEM$ determination, the recoil measurement of Ref.~\cite{Morel:2020dww} was replaced by the improved measurement from the same group~\cite{Bouchendira:2011}, and the  measurement of the electron magnetic moment of Ref.~\cite{Hanneke:2008tm} by the recently improved measurement~\cite{Fan:2022eto}. 
Concerning the electron mass, the only change is the recently discovered long-distance contribution of order $\alphaEM^2(Z\alphaEM)^5\ln(Z\alphaEM)$~\cite{Czarnecki:2020} that is included in the theoretical expression for the bound-electron $g$-factor. 

The input data for all of the utilized relative atomic masses were updated to their AME2020 values~\cite{Huang:2021nwk,Wang:2021xhn}. 
In particular, these values take into account the recent high-precision measurements of the proton, deuteron and HD$^+$ masses~\cite{Heisse:2019xnz,Rau:2022xnn} and the deuteron--proton mass ratio~\cite{Fink:2020}. 
Note that the latest measurement of $m_p/m_d$~\cite{Fink:2021yfd} is not included in the AME2020. 
We chose not to add it separately because of the possible (unknown to us) correlations with the AME2020 values of the hydrogen and deuterium atomic masses.

Moreover, we introduced measurements of transition frequencies in simple molecular or molecule-like systems: 
the hydrogen deuteride molecular ion (\hdplus)~\cite{Alighanbari:2020,Patra:2020,Kortunov:2021rfe}, and the antiprotonic helium atom (\hethreepbar~and  \hefourpbar)~\cite{Hori:2011,Hori:2016}.  
The main merit of adding these systems is their large sensitivity to the electron--nucleus mass ratios, enhanced by about three orders of magnitude relative to hydrogen.  
Due to this, the uncertainty on $m_e/m_p$ is currently the largest contribution to the uncertainty of the SM prediction~\cite{Korobov:2017tvt,Korobov:2021}. 
On the other hand, this then allows for improved determinations of the $m_e/m_p$ mass ratio from comparisons of theory and experiment~\cite{Alighanbari:2020,Patra:2020,Kortunov:2021rfe,Korobov:2021}. 
These determinations are in good agreement with the latest Penning trap measurements~\cite{Huang:2021nwk,Wang:2021xhn,Heisse:2019xnz,Rau:2022xnn,Fink:2020} and have similar uncertainties, in the $10^{-11}$ range. 
They constitute a test of the SM, which, despite being less precise than that obtained from hydrogen-like atoms, is much more sensitive to the NP models in which the mediators have increased couplings to nuclei. 
Examples of such models are the Higgs portal and the hadrophilic scalar, introduced in Section \ref{sec:NPModels} in the main text. 
This feature has been exploited to constrain Yukawa-type forces between hadrons~\cite{Salumbides:2013aga,Germann:2021koc}, and is the major benefit of including the \hdplus data in the DATA22 dataset. 
Similarly, the \hepbar spectroscopy leads to the determination of the antiproton-to-electron mass ratio (or, equivalently, the $m_p/m_e$ ratio, assuming CPT symmetry) with an uncertainty slightly below $10^{-9}$~\cite{Hori:2016}. 
Although the achieved precision is lower, these data provide useful constraints on NP models for higher mediator masses relative to the HD$^+$ constraints, due to the smaller average distance between the nuclei~\cite{Germann:2021koc}. 
The \hepbar energy levels depend on additional parameters, \ie, the masses and charge radii of the $\alpha$ particle and the helion. 
Their masses are deduced from the AME 2020 values of helium-3 and helium-4 masses and their theoretical binding energies. 
The charge radii are determined from the muonic helium spectroscopy data~\cite{Krauth:2021,Krauth:2017}, which we therefore added to the DATA22 dataset.

%%%%%%%%%%%%%%%%%%%%%%%%%%%%%%%%%%%%%%%%%%%%%%%%%%%%%%%%%%%%%%%%%%%%%%
%\section{CODATA18 dataset}
%\label{sec:CODATA2018dataset}
%%%%%%%%%%%%%%%%%%%%%%%%%%%%%%%%%%%%%%%%%%%%%%%%%%%%%%%%%%%%%%%%%%%%%%

%%%%%%%%%%%%%%%%%%%%%%%%%%%%%%%%%%%%%%%%%%%%%%%%%%%%%%%%%%%%%%%%%%%%%%
%\section{DATA22 dataset}
%\label{sec:postCODATA2018dataset}
%%%%%%%%%%%%%%%%%%%%%%%%%%%%%%%%%%%%%%%%%%%%%%%%%%%%%%%%%%%%%%%%%%%%%%

%%%%%%%%%%%%%%%%%%%%%%%%%%%%%%%%%%%%%%%%%%%%%%%%%%%%%%%%%%%%%%%%%%%%%%
\begin{table}[t]
    \centering
    \renewcommand{\arraystretch}{1.2}
    \begin{tabular}{cccc c}
    \toprule\hline
Label & \multicolumn{1}{c}{Input datum} & \multicolumn{1}{c}{Value} & \multicolumn{1}{c}{Rel. uncert.} & Reference \\
\midrule
A30 & $\nuh(1S_{1/2}-3S_{1/2})$     & \num{2922743278665.79(72)} kHz & \num{2.5e-13}  & Grinin {\it et al.}\,\cite{Grinin:2020}\\
A31 & $\nuh(2S_{1/2}-8D_{5/2})$     & \num{770649561570.9}(2.0) kHz & \num{2.6e-12}  & Brandt {\it et al.}\,\cite{Brandt:2021yor}\\
D1 & $ a_{e} \equiv \frac{1}{2}(g-2)_e$     & \num{1.159 652 180 59(13)} $\times 10^{-3}$& \num{1.1e-10} & Fan {\it et al.}\,\cite{Fan:2022eto}\\ 
D3 & $h/m_{\text{Rb}} ({}^{87}\text{Rb}) $ & \num{4.59135925890(65)} $\times 10^{-9} \,\text{m}^2\text{s}^{-1}$ & \num{1.4e-10} &  Morel {\it et al.}\,\cite{Morel:2020dww}\\
D5 & $ A_{\rm r} ({}^{87}\text{Rb})$ & \num{86.909180529(6)} & \num{6.9e-11} &  AME 2020 \cite{Wang:2021xhn}\\
D6 & $A_{\rm r} ({}^{133}\text{Cs}) $ & \num{132.905451958(8)} & \num{6.0e-11} & AME 2020 \cite{Wang:2021xhn} \\
D9 & $\delta_\text{C}$ & $0.0(9.4)$ $\times 10^{-12}$ & \num{4.9e-12} & Czarnecki {\it et al.}\,\cite{Czarnecki:2020} \\
D13 & $\delta_\text{Si}$ & $0.0(5.8)$ $\times 10^{-10}$ & \num{2.8e-10} & Czarnecki {\it et al.}\,\cite{Czarnecki:2020}\\
D11 & $A_{\rm r} ({}^{28}\text{Si}) $ & \num{27.97692653442(55)} & \num{2.0e-11} & AME 2020 \cite{Wang:2021xhn}   \\
D14 & $ A_{\rm r} ({}^{2}\text{H}) $     & \num{2.014101777844(15)} & \num{7.4e-12} & AME 2020 \cite{Wang:2021xhn}  \\
D15 & $ \Delta E_{\rm B}  ({}^{2}\text{H}^{+}) /hc$     & \num{1.0970861455299(10)} $\times 10^{7}\, \text{m}^{-1}$ & \num{9.1e-13}  & NIST ASD 2021 \cite{NIST_ASD}\\
D19 & $A_{\rm r} ({}^{1}\text{H}) $ & \num{1.007825031898(14)} & \num{1.4e-11} & AME 2020 \cite{Wang:2021xhn}   \\
\OLD{D23} & \OLD{$ \Delta E_{\rm B}  ({}^{12}\text{C}^{6+}) /hc$} & $---$ & $---$ & $---$ \\
E1 & $ \nu_{\text{HD}^+} ((0,0) - (0,1))$     & \num{1 314 925 752.910(17)} kHz & \num{1.3e-11}  &Alighanbari {\it et al.}\,\cite{Alighanbari:2020}\\
E2 & $\nu_{\text{HD}^+} ((0,0) - (1,1))$     & \num{58 605 052 164.24(86)} kHz & \num{1.5e-11}  & Kortunov {\it et al.}\,\cite{Kortunov:2021rfe} \\
E3 & $ \nu_{\text{HD}^+} ((0,3) - (9,3))$     & \num{415 264 925 501.8}(1.3) kHz & \num{3.1e-12}  & Patra {\it et al.}\,\cite{Patra:2020} $+$ Germann {\it et al.}\,\cite{Germann:2021koc} \\
G1 & $ \nu_{\bar{\text{p}}^4\text{He}} ((32,31) - (31,30))$ & \num{1 132 609 226.7}(4.0) MHz & \num{3.5e-9} & Hori {\it et al.}\,\cite{Hori:2016} \\
G2 & $ \nu_{\bar{\text{p}}^4\text{He}} ((33,32) - (31,30))$ & \num{2 145 054 858(7)} MHz & \num{3.4e-9} & Hori {\it et al.}\,\cite{Hori:2011} \\
G3 & $ \nu_{\bar{\text{p}}^3\text{He}} ((32,31) - (31,30))$ & \num{1 043 128 581(6)} MHz & \num{6.2e-9} & Hori {\it et al.}\,\cite{Hori:2016} \\
G4 & $ \nu_{\bar{\text{p}}^3\text{He}} ((35,33) - (33,31))$ & \num{1 553 643 100(10)} MHz & \num{6.7e-9} & Hori {\it et al.}\,\cite{Hori:2011} \\
I1 & $ E_{\rm LS}(\mu^4\text{He})$ & \num{1378.521(48)} meV & \num{3.5e-5} & Krauth {\it et al.}\,\cite{Krauth:2021} \\
I2 & $ E_{\rm LS}(\mu^3\text{He})$ & \num{1258.586(49)} meV & \num{3.9e-5} & Krauth\,\cite{Krauth:2017}\\
\hline \bottomrule
    \end{tabular}
    \caption{The new inputs for the DATA22 dataset. $\nu_{\text{HD}^+}((v, L) - (v',L'))$ corresponds to the spin-averaged frequency of the transition between rovibrational states $(v,L)$ and $(v',L')$ of \hdplus, where $v$ ($v'$) denote the initial (final) vibrational state, and $L$ ($L'$) the initial (final) orbital angular momentum quantum number. 
    Similarly, $\nu_{\bar{\text{p}}\text{He}} ((n,l) - (n',l'))$ is the frequency of the transition between states $(n,l)$ and $(n',l')$ of \hepbar, where $n$ ($n'$) and $l$ ($l'$) respectively denote the principal and orbital quantum numbers of the antiproton in the initial (final) state. The ``$---$'' symbol indicates that the input datum has been removed from the CODATA18 dataset.} 
    \label{tab:inputsNEW}
\end{table}
%%%%%%%%%%%%%%%%%%%%%%%%%%%%%%%%%%%%%%%%%%%%%%%%%%%%%%%%%%%%%%%%%%%%%%

%%%%%%%%%%%%%%%%%%%%%%%%%%%%%%%%%%%%%%%%%%%%%%%%%%%%%%%%%%%%%%%%%%%%%%
\begin{table}[t]
    \centering
    \renewcommand{\arraystretch}{1.2}
    \begin{tabular}{cccc}
    \toprule\hline
Label & \multicolumn{1}{c}{Input datum} & \multicolumn{1}{c}{Value (kHz)} & \multicolumn{1}{c}{Rel. uncert.} \\
\midrule
B1  & $\dH(1S_{1/2})/h$ & \num{0.0}(\num{1.3}) & \num{4.0e-13} \\
B2  & $\dH(2S_{1/2})/h$ & \num{0.00(16)}   & \num{2.0e-13} \\
B3  & $\dH(3S_{1/2})/h$ & \num{0.000(49)}  & \num{1.3e-13} \\
B4  & $\dH(4S_{1/2})/h$ & \num{0.000(21)}  & \num{1.0e-13} \\
B5  & $\dH(6S_{1/2})/h$ & \num{0.000(11)}  & \num{1.2e-13} \\
B6  & $\dH(8S_{1/2})/h$ & \num{0.0000(47)} & \num{9.2e-14} \\
B17 & $\dD(1S_{1/2})/h$ & \num{0.0}(\num{1.2})   & \num{3.5e-13} \\
B18 & $\dD(2S_{1/2})/h$ & \num{0.00(15)}   & \num{1.8e-13} \\
B19 & $\dD(4S_{1/2})/h$ & \num{0.000(18)}  & \num{9.0e-14} \\
B20 & $\dD(8S_{1/2})/h$ & \num{0.0000(46)} & \num{8.9e-14} \\
\hline\bottomrule
    \end{tabular}
    \caption{Updated input data for the additive energy corrections accounting for missing contributions to the theoretical description of H and D energy levels. 
    Uncertainties of non-$S$ energy levels are not given here as they are unchanged with respect to Table~VIII of~\cite{Tiesinga:2021myr}.}
    \label{tab:HDuncertNEW}
\end{table}

\begin{table}[t]
\newcolumntype{R}{>{\raggedleft\arraybackslash}p{97px}}
    \centering
    \renewcommand{\arraystretch}{1.2}
    \begin{tabular}{RRRRR}
    \toprule\hline
    \multicolumn{5}{c}{Updated correlation coefficients in hydrogen and deuterium}\\
    \midrule
$r({\rm B1},{\rm B2}) = 0.9960$ & $r({\rm B1},{\rm B3}) = 0.9948$ & $r({\rm B1},{\rm B4}) = 0.9860$ & $r({\rm B1},{\rm B5}) = 0.5409$ & $r({\rm B1},{\rm B6}) = 0.5393$ \\
$r({\rm B1},{\rm B{17}}) = 0.9421$ & $r({\rm B1},{\rm B{18}}) = 0.9393$ & $r({\rm B1},{\rm B{19}}) = 0.9273$ & $r({\rm B1},{\rm B{20}}) = 0.4648$ & $r({\rm B2},{\rm B3}) = 0.9948$ \\
$r({\rm B2},{\rm B4}) = 0.9860$ & $r({\rm B2},{\rm B5}) = 0.5409$ & $r({\rm B2},{\rm B6}) = 0.5393$ & $r({\rm B2},{\rm B{17}}) = 0.9393$ & $r({\rm B2},{\rm B{18}}) = 0.9421$ \\
$r({\rm B2},{\rm B{19}}) = 0.9273$ & $r({\rm B2},{\rm B{20}}) = 0.4648$ & $r({\rm B3},{\rm B4}) = 0.9848$ & $r({\rm B3},{\rm B5}) = 0.5402$ & $r({\rm B3},{\rm B6}) = 0.5387$ \\
$r({\rm B3},{\rm B{17}}) = 0.9382$ & $r({\rm B3},{\rm B{18}}) = 0.9382$ & $r({\rm B3},{\rm B{19}}) = 0.9261$ & $r({\rm B3},{\rm B{20}}) = 0.4642$ & $r({\rm B4},{\rm B5}) = 0.5354$ \\
$r({\rm B4},{\rm B6}) = 0.5339$ & $r({\rm B4},{\rm B{17}}) = 0.9299$ & $r({\rm B4},{\rm B{18}}) = 0.9299$ & $r({\rm B4},{\rm B{19}}) = 0.9432$ & $r({\rm B4},{\rm B{20}}) = 0.4601$ \\
$r({\rm B5},{\rm B6}) = 0.2929$ & $r({\rm B5},{\rm B{17}}) = 0.5101$ & $r({\rm B5},{\rm B{18}}) = 0.5101$ & $r({\rm B5},{\rm B{19}}) = 0.5035$ & $r({\rm B5},{\rm B{20}}) = 0.2524$ \\
$r({\rm B6},{\rm B{17}}) = 0.5086$ & $r({\rm B6},{\rm B{18}}) = 0.5086$ & $r({\rm B6},{\rm B{19}}) = 0.5021$ & $r({\rm B6},{\rm B{20}}) = 0.9830$ & $r({\rm B7},{\rm B8}) = 0.0001$ \\
$r({\rm B9},{\rm B{10}}) = 0.0001$ & $r({\rm B{11}},{\rm B{12}}) = 0.6738$ & $r({\rm B{11}},{\rm B{21}}) = 0.9428$ & $r({\rm B{11}},{\rm B{22}}) = 0.4797$ & $r({\rm B{12}},{\rm B{21}}) = 0.4781$ \\
$r({\rm B{12}},{\rm B{22}}) = 0.9428$ & $r({\rm B{13}},{\rm B{14}}) = 0.2061$ & $r({\rm B{13}},{\rm B{15}}) = 0.2392$ & $r({\rm B{13}},{\rm B{16}}) = 0.2421$ & $r({\rm B{13}},{\rm B{23}}) = 0.9738$ \\
$r({\rm B{13}},{\rm B{24}}) = 0.1331$ & $r({\rm B{13}},{\rm B{25}}) = 0.1351$ & $r({\rm B{14}},{\rm B{15}}) = 0.2225$ & $r({\rm B{14}},{\rm B{16}}) = 0.2252$ & $r({\rm B{14}},{\rm B{23}}) = 0.1128$ \\
$r({\rm B{14}},{\rm B{24}}) = 0.1238$ & $r({\rm B{14}},{\rm B{25}}) = 0.1257$ & $r({\rm B{15}},{\rm B{16}}) = 0.2613$ & $r({\rm B{15}},{\rm B{23}}) = 0.1309$ & $r({\rm B{15}},{\rm B{24}}) = 0.9698$ \\
$r({\rm B{15}},{\rm B{25}}) = 0.1459$ & $r({\rm B{16}},{\rm B{23}}) = 0.1325$ & $r({\rm B{16}},{\rm B{24}}) = 0.1455$ & $r({\rm B{16}},{\rm B{25}}) = 0.9692$ & $r({\rm B{17}},{\rm B{18}}) = 0.9979$ \\
$r({\rm B{17}},{\rm B{19}}) = 0.9851$ & $r({\rm B{17}},{\rm B{20}}) = 0.4938$ & $r({\rm B{18}},{\rm B{19}}) = 0.9851$ & $r({\rm B{18}},{\rm B{20}}) = 0.4938$ & $r({\rm B{19}},{\rm B{20}}) = 0.4874$ \\
$r({\rm B{21}},{\rm B{22}}) = 0.3404$ & $r({\rm B{23}},{\rm B{24}}) = 0.0729$ & $r({\rm B{23}},{\rm B{25}}) = 0.0740$ & $r({\rm B{24}},{\rm B{25}}) = 0.0812$ & \\
\hline\bottomrule
    \end{tabular}
    \caption{Updated correlation coefficients $r(\text{Bi},\text{Bj}) \geq 0.0001$ between input data for the hydrogen and deuterium energy corrections. 
    }
    \label{tab:HDcorrelNEW}
\end{table}
%%%%%%%%%%%%%%%%%%%%%%%%%%%%%%%%%%%%%%%%%%%%%%%%%%%%%%%%%%%%%%%%%%%%%%

%%%%%%%%%%%%%%%%%%%%%%%%%%%%%%%%%%%%%%%%%%%%%%%%%%%%%%%%%%%%%%%%%%%%%%
\begin{table}[h]
    \centering
    \renewcommand{\arraystretch}{1.2}
    \begin{tabular}{cccc}
    \toprule\hline
Label & \multicolumn{1}{c}{Input datum} & \multicolumn{1}{c}{Value (kHz)} & \multicolumn{1}{c}{Rel. uncert.} \\
\midrule
F1  & $\dHDplus(0,0)/h$ & \num{0}(\num{21.28}) & \num{5.4e-12} \\
F2  & $\dHDplus(0,1)/h$ & \num{0}(\num{21.27}) & \num{5.4e-12} \\
F3  & $\dHDplus(0,3)/h$ & \num{0}(\num{21.17}) & \num{5.4e-12} \\
F4  & $\dHDplus(1,1)/h$ & \num{0}(\num{20.80}) & \num{5.4e-12} \\
F5  & $\dHDplus(9,3)/h$ & \num{0}(\num{18.18}) & \num{5.2e-12} \\
H1  & $\dHefourpbar(31,30)/h$ & \num{0(1454)} & \num{6.0e-11} \\
H2  & $\dHefourpbar(32,31)/h$ & \num{0(1514)} & \num{6.6e-11} \\
H3  & $\dHefourpbar(33,32)/h$ & \num{0(1595)} & \num{7.2e-11} \\
H4  & $\dHethreepbar(31,30)/h$ & \num{0(1595)} & \num{6.9e-11} \\
H5 & $\dHethreepbar(32,31)/h$ & \num{0(1625)} & \num{7.4e-11} \\
H6 & $\dHethreepbar(33,31)/h$ & \num{0(1804)} & \num{8.5e-11} \\
H7 & $\dHethreepbar(35,33)/h$ & \num{0(1998)} & \num{1.0e-10} \\
I3 & $\delta E_{\rm LS}(\mu^4{\rm He})$ & \num{0.000(293)}\,meV & \num{2.1e-4} \\
I4 & $\delta E_{\rm LS}(\mu^3{\rm He})$& \num{0.000(521)}\,meV & \num{4.1e-4} \\
\hline\bottomrule
    \end{tabular}
    \caption{Input data for the additive energy corrections accounting for missing contributions to the theoretical description of energy levels of three-body systems, \hdplus (F) and \hepbar (H), and the Lamb shift of muonic helium ions (I).}
    \label{tab:3buncert}
\end{table}

\begin{table}[h]
\newcolumntype{R}{>{\raggedleft\arraybackslash}p{97px}}
    \centering
    \renewcommand{\arraystretch}{1.2}
    \begin{tabular}{RRRRR} 
    \toprule\hline
    \multicolumn{5}{c}{Correlation coefficients in \hdplus and \hepbar}\\
    \midrule
$r({\rm F1},{\rm F2}) = 1.0000$ & $r({\rm F1},{\rm F3}) = 1.0000$ & $r({\rm F1},{\rm F4}) = 1.0000$ & $r({\rm F1},{\rm F5}) = 0.9980$ & $r({\rm F2},{\rm F3}) = 1.0000$ \\
$r({\rm F2},{\rm F4}) = 1.0000$ & $r({\rm F2},{\rm F5}) = 0.9980$ & $r({\rm F3},{\rm F4}) = 1.0000$ & $r({\rm F3},{\rm F5}) = 0.9980$ & $r({\rm F4},{\rm F5}) = 0.9982$ \\
$r({\rm H1},{\rm H2}) = 0.9950$ & $r({\rm H1},{\rm H3}) = 0.9893$ & $r({\rm H1},{\rm H4}) = 0.9683$ & $r({\rm H1},{\rm H5}) = 0.9941$ & $r({\rm H1},{\rm H6}) = 0.9771$ \\
$r({\rm H1},{\rm H7}) = 0.9744$ & $r({\rm H2},{\rm H3}) = 0.9980$ & $r({\rm H2},{\rm H4}) = 0.9693$ & $r({\rm H2},{\rm H5}) = 0.9996$ & $r({\rm H2},{\rm H6}) = 0.9855$ \\
$r({\rm H2},{\rm H7}) = 0.9891$ & $r({\rm H3},{\rm H4}) = 0.9646$ & $r({\rm H3},{\rm H5}) = 0.9984$ & $r({\rm H3},{\rm H6}) = 0.9868$ & $r({\rm H3},{\rm H7}) = 0.9957$ \\
$r({\rm H4},{\rm H5}) = 0.9686$ & $r({\rm H4},{\rm H6}) = 0.9526$ & $r({\rm H4},{\rm H7}) = 0.9513$ & $r({\rm H5},{\rm H6}) = 0.9859$ & $r({\rm H5},{\rm H7}) = 0.9906$ \\
$r({\rm H6},{\rm H7}) = 0.9830$ & & & & \\
\hline\bottomrule
    \end{tabular}
    \caption{Correlation coefficients $r(\text{Fi},\text{Fj}) \geq 0.0001$ and $r(\text{Hi},\text{Hj}) \geq 0.0001$ between input data for the \hdplus and \hepbar~energy corrections, respectively.}
    \label{tab:3bcorrel}
\end{table}
%%%%%%%%%%%%%%%%%%%%%%%%%%%%%%%%%%%%%%%%%%%%%%%%%%%%%%%%%%%%%%%%%%%%%%

The new inputs for the DATA22 dataset are listed in Table~\ref{tab:inputsNEW}.
The updates of theoretical uncertainties for hydrogen and deuterium levels, based on~Refs.~\cite{Yerokhin:2019,Karshenboim:2019a,Szafron:2019,Karshenboim:2019b}, and the correlation matrix (calculated following the method of Ref.~\cite{Tiesinga:2021myr}) are collected in Tables~\ref{tab:HDuncertNEW} and~\ref{tab:HDcorrelNEW}, respectively.
Moreover, an expansion factor of $2.4$ is applied for the fine-structure constant data, \ie, datapoints D3 
and D4, because in this case the discrepancies cannot be due to NP, and are most likely due to overlooked or underestimated systematics. 
Note that in the DATA22 inputs, we do not include newer input values for $r_p$ from $e$-$p$ scattering.
These include a new result of the PRad collaboration, $r_p =\num{0.831(14)}$\,fm~\cite{Xiong:2019umf}, and a reanalysis of modern data, giving $r_p =\num{0.847(8)}$\,fm~\cite{Cui:2021vgm}.
In both cases, the determination of the proton charge radius from the electron-proton scattering data is not precise enough to make an appreciable difference in the fit.

The SM calculations of energy levels in the three-body systems \hdplus and \hepbar were carried 
out using the nonrelativistic QED~(NRQED) approach. Energy levels are expanded in powers of $\alphaEM$: 
\begin{align}
    \label{eq:Etot}
    E 
=   E^{(0)} + E^{(2)} + E^{(3)} 
    + E^{(4)} + E^{(5)} + E^{(6+)} 
    + E_{\rm nuc} + E_{\rm other}  \,.
\end{align}
Here $E^{(n)}$ is the total contribution of order $\Rinf \alpha^n$, $E_{\rm nuc}$ corresponds to nuclear structure (finite size and polarizability) corrections, and $E_{\rm other}$ to the muonic and hadronic vacuum polarization corrections.
The first term, $E^{(0)}$, is the nonrelativistic energy. 
It was calculated by solving the three-body Schr\"odinger equation with high accuracy using a variational method~\cite{Korobov:2000,Schiller:2005}, in conjunction with the coordinate rotation~(CCR) method for the case of resonant (quasibound) states in antiprotonic helium~\cite{Korobov:2014hep}.
The leading relativistic correction, $E^{(2)}$, described by the Breit-Pauli Hamiltonian, was calculated in~\cite{Aznabayev:2019} for \hdplus rovibrational states and in~\cite{Korobov:2014hep,Korobov:2003} for \hepbar\!\!.
The $E^{(3)}$ term gives the leading radiative correction. 
It involves a numerically challenging quantity, the Bethe logarithm, which was obtained with high precision in~\cite{Korobov:2012} for \hdplus and in~\cite{Korobov:2014hep} for \hepbar resonant states. 
Other relevant numerical data for this correction can be found in~\cite{Aznabayev:2019,Korobov:2014hep}.
The $\Rinf \alphaEM^4$-order correction, $E^{(4)}$, is made up of several contributions. 
One- and two-loop radiative corrections are given in~\cite{Korobov:2006,Korobov:2008}. 
The relativistic correction was calculated in~\cite{Korobov:2007}, and complete numerical data for \hdplus can be found in the Supplemental material of~\cite{Korobov:2021}. 
The relativistic-recoil contribution was estimated from hydrogenlike atom theory~\cite{Pachucki:1997}, and the radiative-recoil term was taken from~\cite{Pachucki:1995,Czarnecki:2001}.

It is worth noting that the relativistic correction, as well as several higher order corrections (the $\Rinf \alpha^5$ and $\Rinf \alpha^6$ terms), were calculated
in the adiabatic approximation, where the wave function is written as a product of electronic and nuclear (rovibrational) wavefunctions. 
Electronic wavefunctions were obtained by solving the Schr\"odinger equation in the field of two clamped nuclei using a variational method~\cite{Tsogbayar:2006}. 
In the first step, the $\Rinf \alpha^4$ relativistic corrections for the bound electron were calculated for a range of internuclear distances~\cite{Korobov:2007}. 
Then, the electronic curves were averaged over the nuclear wavefunction. 
The effect of corrections to the vibrational wavefunction induced by electronic corrections also needs to be taken into account~\cite{Korobov:2017tvt, Korobov:2021}.

The radiative correction $E^{(5)}$ comprises the one-loop self-energy~\cite{Korobov:2014,Korobov:2015} and the vacuum polarization (the Uehling potential)~\cite{Karr:2017}. 
Complete numerical data for these two contributions are available in the Supplemental Material of Ref.~\cite{Korobov:2021} for \hdplus. 
The Wichmann-Kroll vacuum polarization term, as well as two-loop and three-loop corrections are also included~\cite{Korobov:2014}. 
It should be noted that in the case of antiprotonic helium, the Uehling correction has only been estimated from the hydrogen atom theory~\cite{Korobov:2015}.

One- and two-loop radiative corrections of order $\Rinf \alpha^6$ have been partially calculated in \hdplus ~in~\cite{Korobov:2017tvt,Korobov:2021} (complete numerical data are available in the Supplemental Material of \cite{Korobov:2021}).  
In antiprotonic helium, they have only been estimated using the hydrogen atom theory~\cite{Korobov:2015}.
In the nuclear finite-size and structure correction $E_{\rm nuc}$, we include the leading-order finite-size correction~\cite{Korobov:2006,Korobov:2008}. 
In \hdplus, we take into account the higher-order nuclear corrections for the deuteron as described in~\cite{Korobov:2021}. 
In \hepbar, we also include these corrections for the helium-3 and helium-4 nuclei using the theoretical expressions presented in~\cite{Yerokhin:2019}. 
Similar corrections for the proton or antiproton are negligibly small at the present level of experimental and theoretical accuracy.
The last term of Eq.~(\ref{eq:Etot}), denoted by $E_{\rm other}$, corresponds to the muonic and hadronic vacuum polarization contributions, with the explicit expressions
given in~\cite{Korobov:2021}.

Theoretical uncertainties, Table~\ref{tab:3buncert}, and their correlations, Table~\ref{tab:3bcorrel}, were estimated following an  approach similar to~\cite{Tiesinga:2021myr}. 
In \hdplus, by far the largest sources of uncertainty are the yet uncalculated nonlogarithmic contributions of order $\Rinf \alpha^6$ in the one-loop self-energy and in the two-loop radiative correction~\cite{Korobov:2017tvt,Korobov:2021}. 
Smaller uncertainties come from the use of the adiabatic approximation in the calculation of some of the high-order correction terms ($\Rinf \alpha^4$ to $\Rinf \alpha^6$), from the relativistic-recoil correction of order $\Rinf \alpha^4 m_e/m_p$, and from the deuteron finite-size correction of order $\Rinf \alpha^4$. 
All these uncertainties are assumed to be fully correlated (``type $u_0$'' uncertainties in the terminology of Ref.~\cite{Tiesinga:2021myr}), leading to correlation coefficients very close to 1. 
In \hepbar\!\!, additional sources of uncertainty come into play. 
Firstly, there is a larger uncertainty associated with one- and two-loop $\Rinf \alpha^6$-order corrections, since these terms have only been estimated from hydrogen atom results. 
Secondly, the one-loop vacuum polarization term of order $\Rinf \alpha^5$ has a significant uncertainty, for the same reason. 
Finally, some of the operator expectation values involved in the calculation of $E^{(2)}$ and $E^{(3)}$ have non-negligible numerical uncertainties. 
These numerical uncertainties are assumed to be uncorrelated (``type $u_n$''~\cite{Tiesinga:2021myr}), leading to smaller correlation coefficients with respect to \hdplus.

The sensitivity of theoretical energy levels to the SM parameters ($\Rinf$, $\alpha$, and nuclear charge radii) is easily obtained by differentiating Eq.~\eqref{eq:Etot}, with the exception of particle masses. 
The dependence on particle masses mainly comes from the nonrelativistic energy (first term of Eq.~\eqref{eq:Etot}), and is obtained numerically. 
The corresponding sensitivity coefficients are thus calculated numerically using the approach outlined in~\cite{Schiller:2005}.

 The values of $g_{\rm SM}$ determined from the fit to DATA22 data under the hypothesis of just SM, {\it i.e.}, no new physics contributions, are listed in Table \ref{tab:CODATA18adjus}. We reiterate that in this adjustment the experimental errors were not increased by expansion factors, unlike in the CODATA18 dataset, except for the input data D3 
and D4. The extraction of SM and NP parameters in the case of new physics is discussed in the main text and in Sections \ref{sec:NP:vs:SM} and \ref{sec:app:extraction:NP}. 

\section{Further details on NP benchmark models}
\label{sec:further:NP}

In the main text   we introduced  in Section \ref{sec:NPModels} five NP benchmark models that were then used to illustrate simultaneous extraction of SM and NP parameters, highlighting relevance of different spectroscopic data. Dark photon, $B-L$ gauge boson, and light Higgs-mixed scalar are all models that were already widely discussed in the literature. Hadrophilic scalar is a straight-forward modification of light Higgs-mixed scalar, taking the couplings in the leptonic sector to vanish. Below we give further details on the more involved modification, the ULD scalar, and also introduce another NP benchmark model, the scalar photon model, which was omitted from the discussion in the main text for brevity. 

\subsection{Up-Lepto-Darko-Philic~(ULD) scalar}\label{sec:app-ULD}

At low energies the ULD benchmark model comprises the SM Higgs and an additional light scalar singlet $\phi$. The light scalar couples to up quarks, electrons and muons, and to a dark sector (SM singlet) fermion $\chi$. For ease of comparison we use the notation for the couplings that is reminiscent of the Higgs-mixed scalar benchmark model, {\it i.e.}, 
\begin{equation}
       \mathcal{L}_\phi= k\frac{m_\ell}{v}\phi \bar\ell \ell+ k\frac{m_u}{v}\phi\bar u u + y_\chi\phi\bar\chi\chi\,, 
\end{equation}
with $\ell=e,\mu$ and $k$ a model-dependent constant, however, the underlying high scale model is different. 
The couplings of $\phi$ to leptons and nucleons are then given by, 
\begin{equation}
{\cal L}_{\rm eff}= g_\ell \phi \bar \ell \ell +g_N \phi \bar N N\,,
\end{equation}
where
\begin{equation}
g_\ell =k\frac{m_\ell}{v}\,, \qquad g_N=k\frac{\kappa_N'  m_N}{v}\,,
\end{equation}
with  $\kappa_p'\simeq0.018(5)$
 and  $\kappa_n'\simeq 0.016(5)$. 
 Here and for the scalar mixed portal couplings we use $\sigma_{u,d}^{p,n}$ values from~\cite{Bishara:2017pfq} that were obtained following the procedure in~\cite{Crivellin:2013ipa}, from $\sigma_{\pi N}=(50\pm15)$ MeV with conservative errors~\cite{Bishara:2017pfq} which
covers the spread between lattice QCD and pionic
atom determinations. For the scalar
operator of the strange quark we use the FLAG value
$\sigma_s^p=\sigma_s^n=(52.9\pm 7.0){\rm~MeV}$~\cite{FlavourLatticeAveragingGroup:2019iem},
obtained by averaging the $N_f=2+1$ lattice QCD
results~\cite{Yang:2015uis, Freeman:2012ry, Durr:2015dna,
  Junnarkar:2013ac, Durr:2011mp} (see also \cite{Alexandrou:2019brg,
  Borsanyi:2020bpd, Yang:2015uis}). 
The NP parameters in the non-relativistic potential \eqref{eq:VNP} are therefore given by
\begin{equation}
q_\ell=\frac{m_\ell}{\sqrt{m_e \kappa_p'm_p}}\,,\qquad  q_N=\frac{\kappa_N'm_N}{\sqrt{m_e \kappa_p'm_p}}\,, \qquad \alpha_\phi=k^2\frac{m_e \kappa_p'm_p}{4\pi v^2}\,.
\end{equation}
Because the electron-proton coupling ratio is enhanced by $\kappa_p/\kappa_p'\sim \cO(20)$  relative to the Higgs portal model, $\phi$ contributions to  electron and muon $g-2$ and electron beam dump signals are significantly enhanced at fixed $\alpha_\phi$. 
Whenever necessary, bounds from beam dump experiments can be evaded assuming $\phi$ decays dominantly to additional invisible particles that could be related to dark matter. The $\phi$ contribution to muon $g-2$ can be reduced in less minimal models that include a custodial symmetry~\cite{Balkin:2021rvh}. In such models the $\phi$ is accompanied by a light pseudoscalar particle which leads to a new spin-dependent force.   
Below we consider two distinct UV realizations of the ULD benchmark model.\\ 

The first possibility is a light scalar that couples to up quarks, electrons and muons through dimension five operators, 
\begin{equation}\label{eq:ULD}
    \mathcal{L}_\phi = \frac{y_\ell'}{\Lambda} \phi \,\bar L_\ell H \ell_R + \frac{y_u'}{\Lambda} \phi \,\bar Q_u \tilde H u_R+{\rm h.c.} \,, 
\end{equation}
where $L_{\ell}$,$Q_u$ and $\ell_R,u_R$ are the SM left-handed doublets and right-handed singlets, respectively, and  $\tilde H = i\sigma_2 H^*$, with $H$ the Higgs doublet. We assume that $y_{\ell,u}'=A\times m_{\ell,u}/v$, with $A$ a universal dimensionless constant. The $\phi$ couplings to electrons, muons and up quarks are then the same as for the SM Higgs, but rescaled by $k=Av/\Lambda$. The favored region of the ULD parameter space is for $|k|\simeq 2.4$, see Fig.~\ref{fig:alphaNP-mNP}. This requires  the $y'$ couplings to be larger than the corresponding SM Yukawa couplings by a factor of $A\simeq 100\times(\Lambda/10\,$TeV). The largest is the coupling to muons, $y_\mu'\simeq 0.04\times  (\Lambda/10\,\text{TeV})$, which remains well perturbative even for a relatively high value of the cutoff scale $\Lambda$.  
The higher dimensional operators could arise from  extra vector-like fermions at the scale $\Lambda$,
\begin{equation}
\mathcal{L}_{\rm UV} = \lambda_\ell\phi \bar \Psi_L^\ell \ell_R + x_\ell\bar L_\ell H \Psi_R^\ell  + \lambda_u \phi \bar\Psi_L^u u_R + x_u \bar Q_u \tilde H \Psi_R^u+{\rm h.c.}\,,
\end{equation}
where $\Psi^{\ell,u}$ are vector-like SU(2)$_L$ singlets of mass $M_{\ell,u}$. Integrating out the $\Psi$'s yields the dimension-five operators in Eq.~\eqref{eq:ULD} with $y_\ell'/\Lambda = x_\ell \lambda_\ell/M_\ell$ and $y_u'/\Lambda = x_u \lambda_u/M_u$.\\

In the second example of a UV model we supplement the SM field content  by an extra Higgs doublet $H'$ 
and a light singlet $S$ that mixes with  $H'$, giving a light mass eigenstate $\phi\simeq S- \sin\theta' H'$, where $\theta'\ll 1$ is the mixing angle. The mixings of the two scalars with the SM Higgs is assumed to be small. The Yukawa couplings of $H'$ to the SM fermions are assumed to be diagonal in the same basis as for the SM Higgs, with the only nonzero values the Yukawa couplings to the electron, the muon and the up quark. 
The SM Higgs $H$ remains the dominant source of the electron, muon and up-quark masses and gives the entirety of the mass to the remaining SM fermions, while a subdominant parts of the electron, muon and up quark masses are due to the $H'$ vacuum expectation value, $\langle H'\rangle=(0, v')/\sqrt{2}$
(for more general models of this type see, {\it e.g.}, \cite{Altmannshofer:2015esa,Botella:2015hoa,Ghosh:2015gpa}). That is, for electron, muon and up quark we have
\begin{equation}
    m_{\ell,u} = m_{\ell,u}^H+m_{\ell,u}^{H'}    \,,
\end{equation}
and $m_i=m_i^H$ for $i= \tau,d,s,c,b,t$. As a simple ansatz we take $m_{\ell,u}^{H'}= \delta\times m_{\ell,u}^H$, with $\delta\ll 1$ a universal factor.
The $\phi$ couplings to electrons, muons and up quarks are then the same as for the SM Higgs, just rescaled by $k= \delta \sin\theta' v/v'$. 
The smallness of $\delta$ guarantees that the SM Higgs boson decays to muons, $h\to\mu^+\mu^-$, remain close to the SM predictions, in agreement with measurements, which require $\delta\lesssim 0.35$ at 90\% CL~\cite{CMS:2020xwi}.
Note that for $|k|\sim\cO(1)$ 
%in this case, while not to much suppressing the $h\to \mu^+\mu^-$ rate ($\delta\ll 1$), 
the $H'$ VEV must be  smaller than the weak scale, $v'/v\simeq \delta \sin\theta'\ll1 $.

\subsection{The scalar photon}

\begin{figure}
    \centering
    \begin{tabular}{cc}
     \includegraphics[width=0.45\columnwidth]{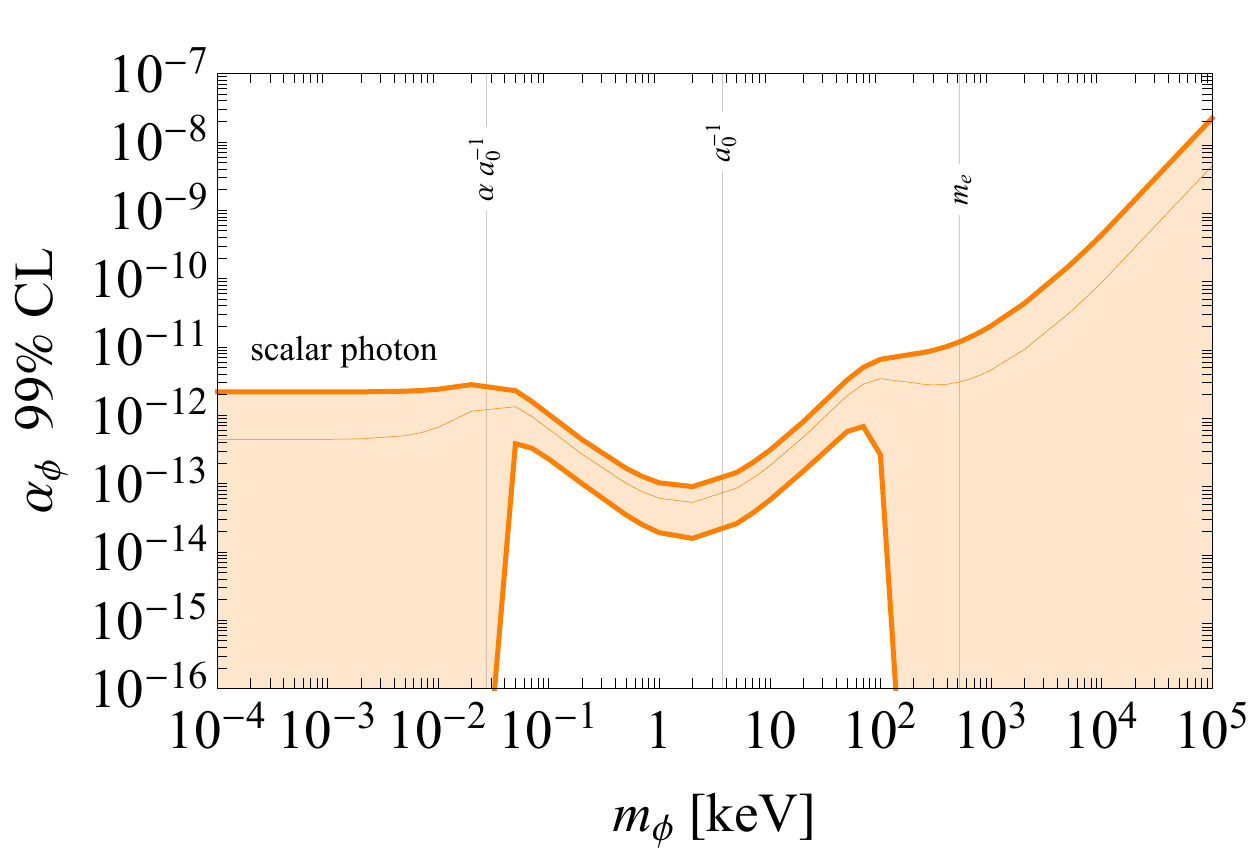}    &  
    \includegraphics[width=0.45\columnwidth]{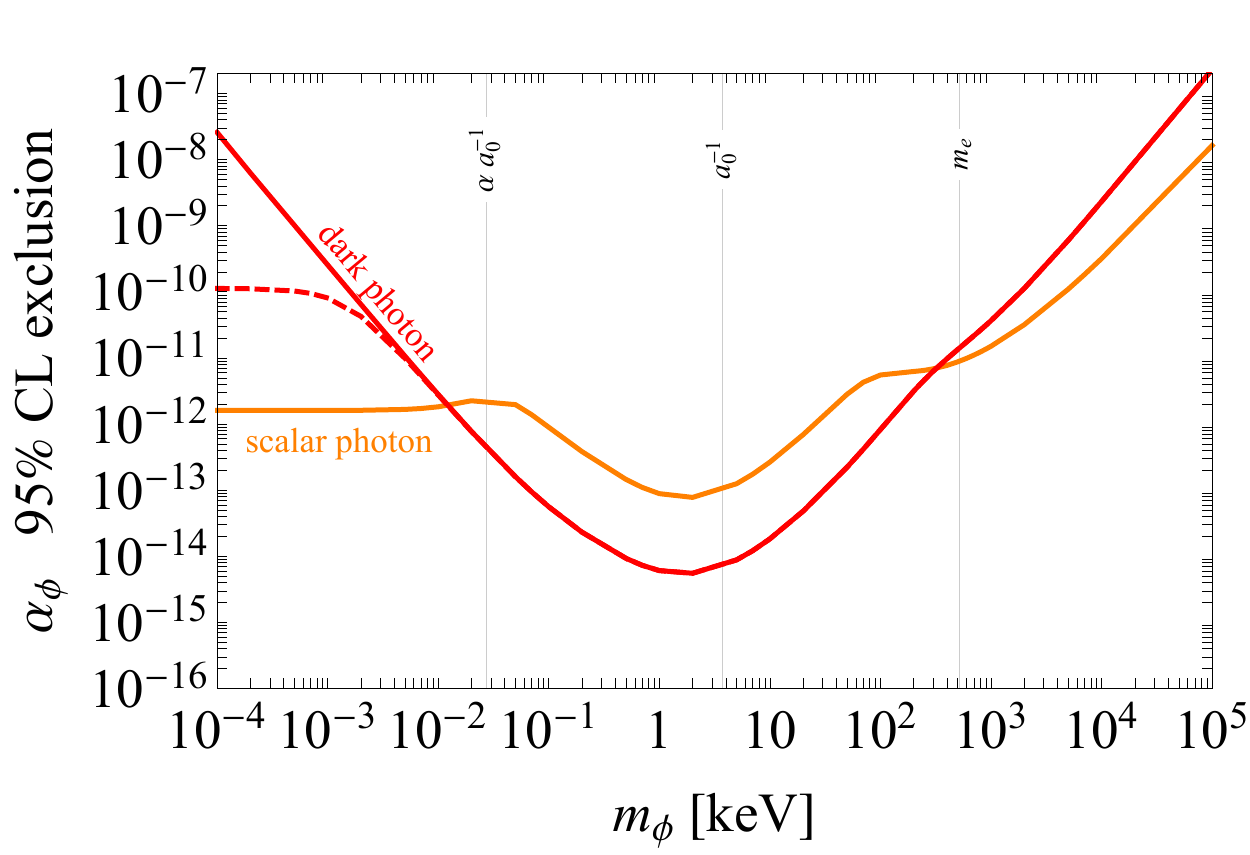}
    \end{tabular}
    \caption{(left) 99\%\,CL interval on the NP coupling $\alphaNP$ as function of $\mNP$ for the scalar photon benchmark model. (right) Comparison of the $95\%$CL exclusion on $\alphaNP$ in the dark photon (red) and scalar photon (orange) models. The dashed line denotes the unphysical bound in the dark photon case that results from the same fit but without redefining  $\alpha$ 
    in order to recover the correct behavior in the long-range ($\mNP\to 0$) regime.}
    \label{fig:DP-vs-SP}
\end{figure}

Finally, we introduce an additional NP benchmark model, the ``scalar photon", which will prove useful in the discussion of the $m_\phi\to 0$ limits in  Section \ref{sec:masslessNP} below.
The model consists of SM and a new light scalar that has the same pattern of couplings to the SM fermions  as the photon, {\it i.e.}, $q_e=q_\mu=-q_p=-1$ and $q_n=0$. That is, the only difference between the dark photon and the scalar photon model is the spin of the mediator. The comparison of bounds for the two models is therefore very informative, and highlights the importance of different subsets of data as well as the importance of implementing the $m_\phi\to 0$ behavior correctly. 
For instance, the right panel of Fig.~\ref{fig:DP-vs-SP} shows that the bounds on $\alpha_\phi$ are rather different in the two models, despite the identical pattern of couplings. 
For $\mNP\gtrsim m_e$, the bound is mostly set by the one-loop NP contribution to the electron $g-2$, which is larger for scalar photon. 
For $ m_\phi a_0\sim \cO(1)$, the sensitivity of the fit to NP is dominated by hydrogen spectroscopy, 
which currently favors an addition of a repulsive NP interaction, thus yielding a stronger bound for dark photon. 
For very light masses $m_\phi \lesssim 1/a_0$, the bounds on $\alpha_\phi$ weaken both for the scalar photon as well as for dark photon, due to the  degeneracy with the QED photon. 
However, while for dark photon the degeneracy with the QED photon is exact in the $m_\phi\to0$ limit, for massless scalar photon the degeneracy is lifted by $\cO(\alpha^2)$ relativistic corrections.  The bound on the scalar photon fine-structure constant therefore flattens below small enough $m_\phi$, for which the NP contributions to the hydrogen energy levels become comparable to the relativistic corrections, {\it i.e.}, for $a_0\mNP \sim \cO(\alpha)$.

\begin{figure}
    \centering
\begin{tabular}{lc}
   \includegraphics[width=0.45\columnwidth]{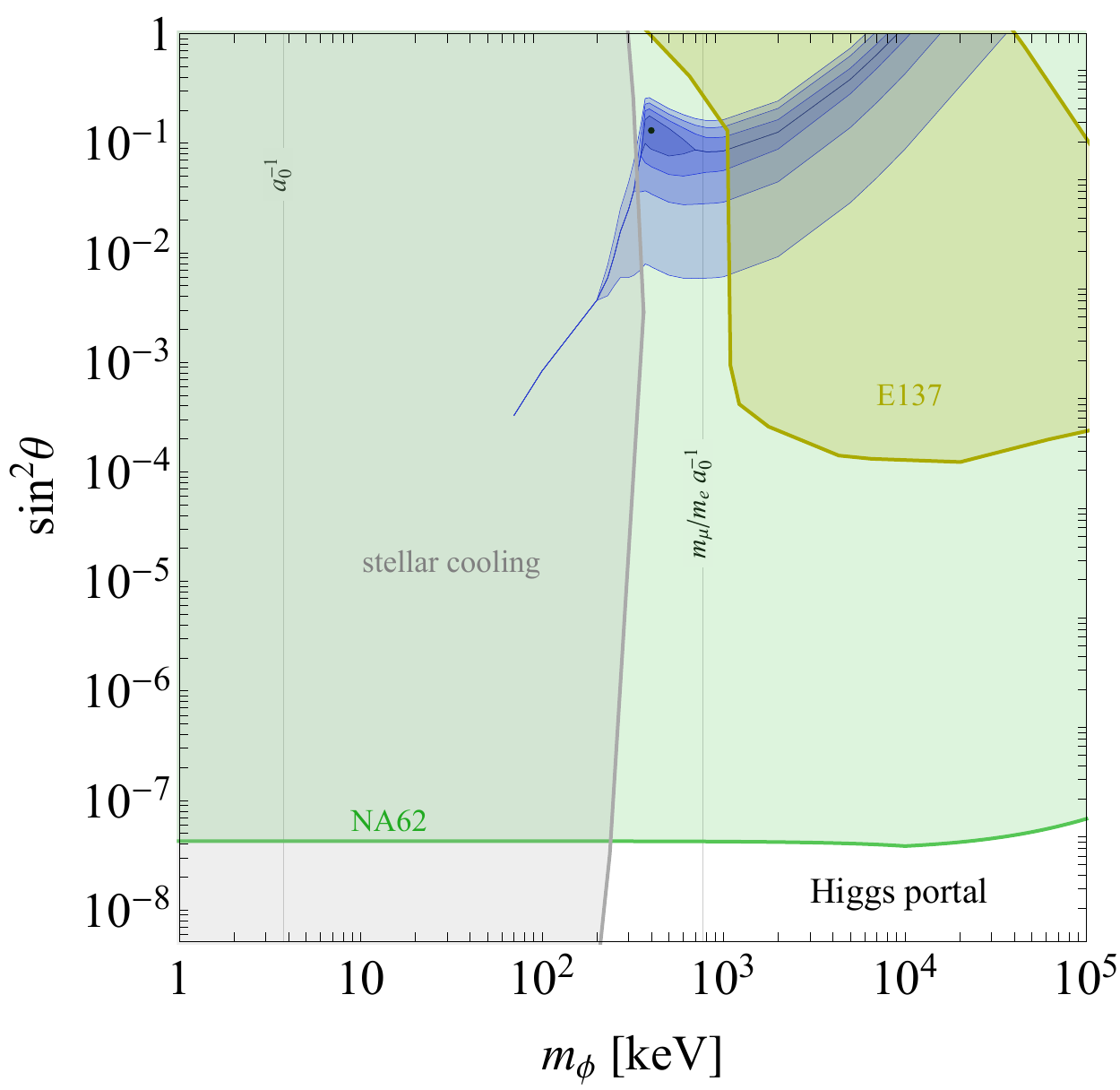} &   \includegraphics[width=0.45\columnwidth]{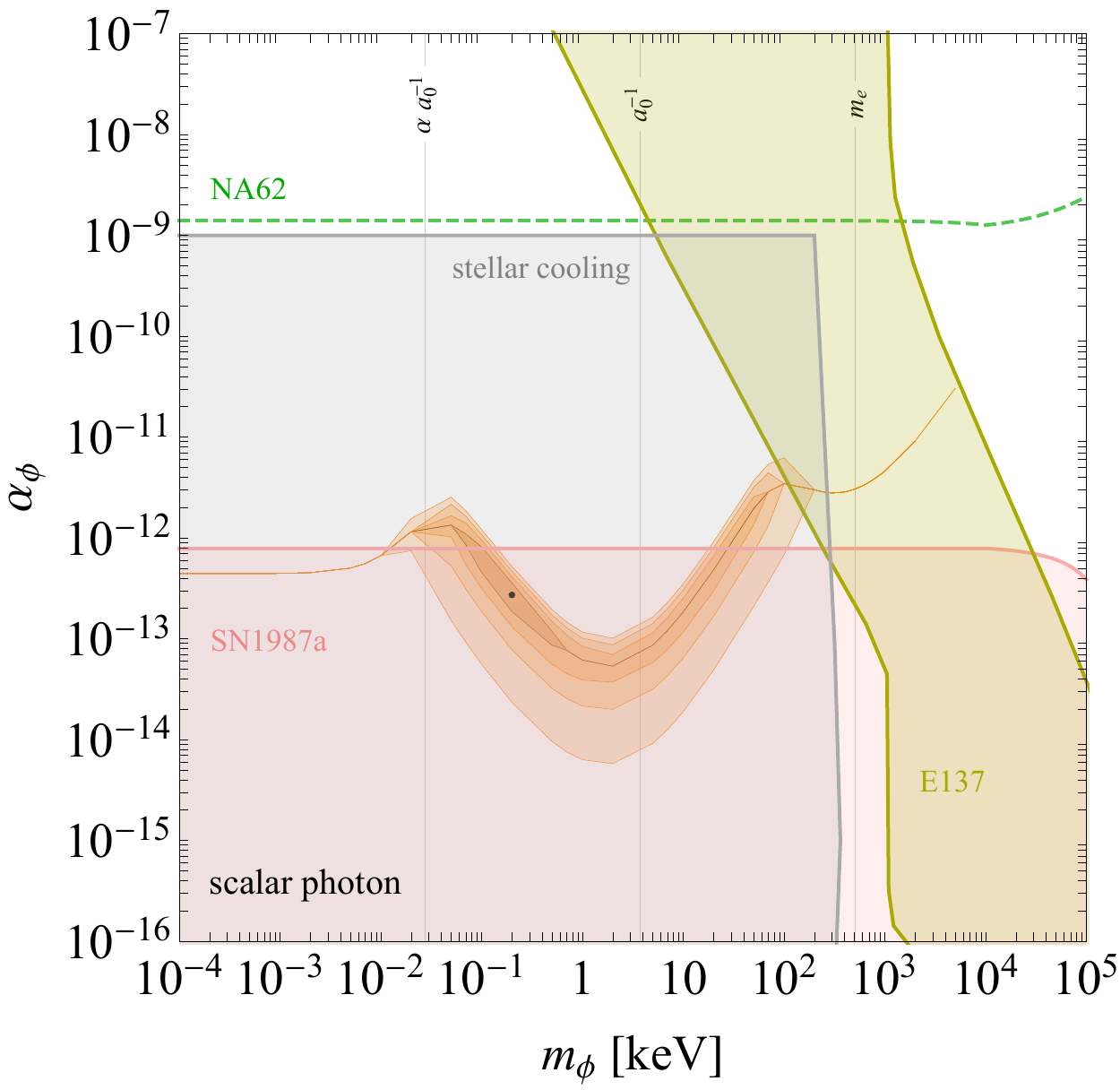} 
\end{tabular}
    \caption{NP parameters favored by atomic spectroscopy in the Higgs portal (left) and the scalar photon (right) benchmark models, profiling over the SM parameters. The black dot indicates the best-fit point and the blue-shaded (left) and orange-shaded (right) areas represent the 1,\,2,\,3,\,4$\sigma$ confidence regions around it. The other shaded areas denote excluded region by SN1987a~\cite{Raffelt:2012sp,Batell:2018fqo,Dev:2020eam} (pink), the $K^+\to\pi^+X$ search at NA62~\cite{NA62:2020xlg} (green, the dashed line is an NNLO estimate), stellar cooling~\cite{Hardy:2016kme} (gray) and the E137 beam dump experiment~\cite{Bjorken:1988as,Liu:2016mqv} (yellow).}
    \label{fig:HP-SPexcess}
\end{figure}

The least-squares adjustment also reveals that the DATA22 dataset favors the scalar photon model over the SM-only hypothesis at the $4.8\sigma$ level, with the best-fit point at $\mNP=0.2\,\keV$ and $\alphaNP=2.7\times 10^{-13}$. The $1\sigma$, $2\sigma$, $3\sigma$, and $4\sigma$ CL contours are shown as orange shaded regions in Fig. \ref{fig:HP-SPexcess} (right panel), along with the exclusions from stellar cooling (grey), SN1987a cooling (red),  NA62 search for $K^+\to \pi^+$+inv (green) and E137 (dark green). The overlap of all the bounds excludes the region favored by the adjustment. Note that even if the SN cooling does not apply \cite{Bar:2019ifz} the favored region is still excluded by the combination of other constraints. 

Figure~\ref{fig:DP-vs-SP} (left panel) shows the allowed range for  $\alpha_\phi$ at 99\% CL assuming a given scalar photon mass (orange shaded region), with the solid line showing the best fit $\alpha_\phi$ for each value of $m_\phi$.  Note that for very light $\phi$, with mass below around $20$\,eV the value $\alpha_\phi=0$ is allowed at 99\%\,CL, {\it i.e.}, the very light scalar photon  is not preferred over the SM. Similarly, the heavy scalar photon, with mass above about 100 keV, is not preferred over the SM, in agreement with Fig. \ref{fig:HP-SPexcess} (right panel).

\section{New physics improvements relative to the SM}
\label{sec:NP:vs:SM}

\begin{figure}
    \centering
    \begin{tabular}{cccc}
      \includegraphics{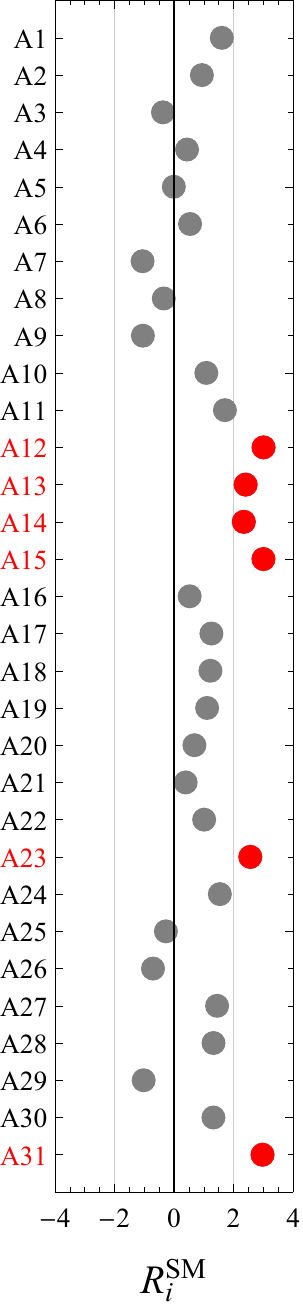}   &     \includegraphics{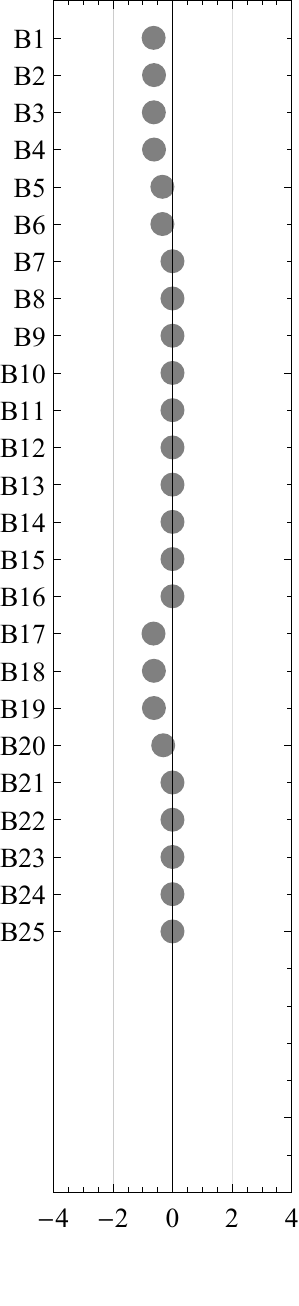} &    \includegraphics{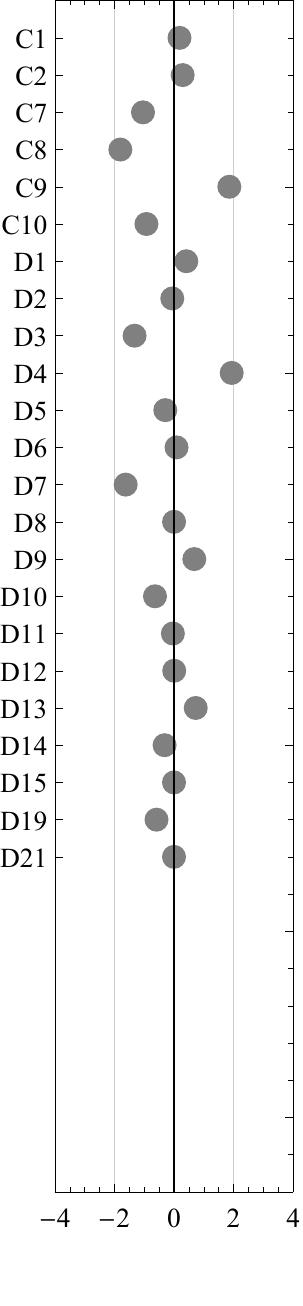} &    \includegraphics{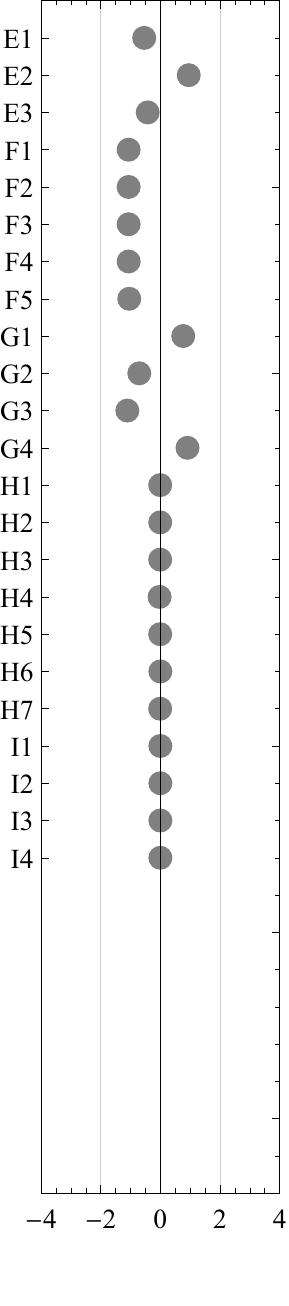}
   \end{tabular}
    \caption{Normalized residuals $R_i^{\rm SM}$, see Eq.~\eqref{eq:Ridef}, of individual input data within the DATA22 dataset.
    Input data satisfying $|R_i^{\rm SM}|\geq 2$ are indicated in red.}
    \label{fig:SMresidu}
\end{figure}

\begin{figure}
    \centering
    \includegraphics[width=0.47\columnwidth]{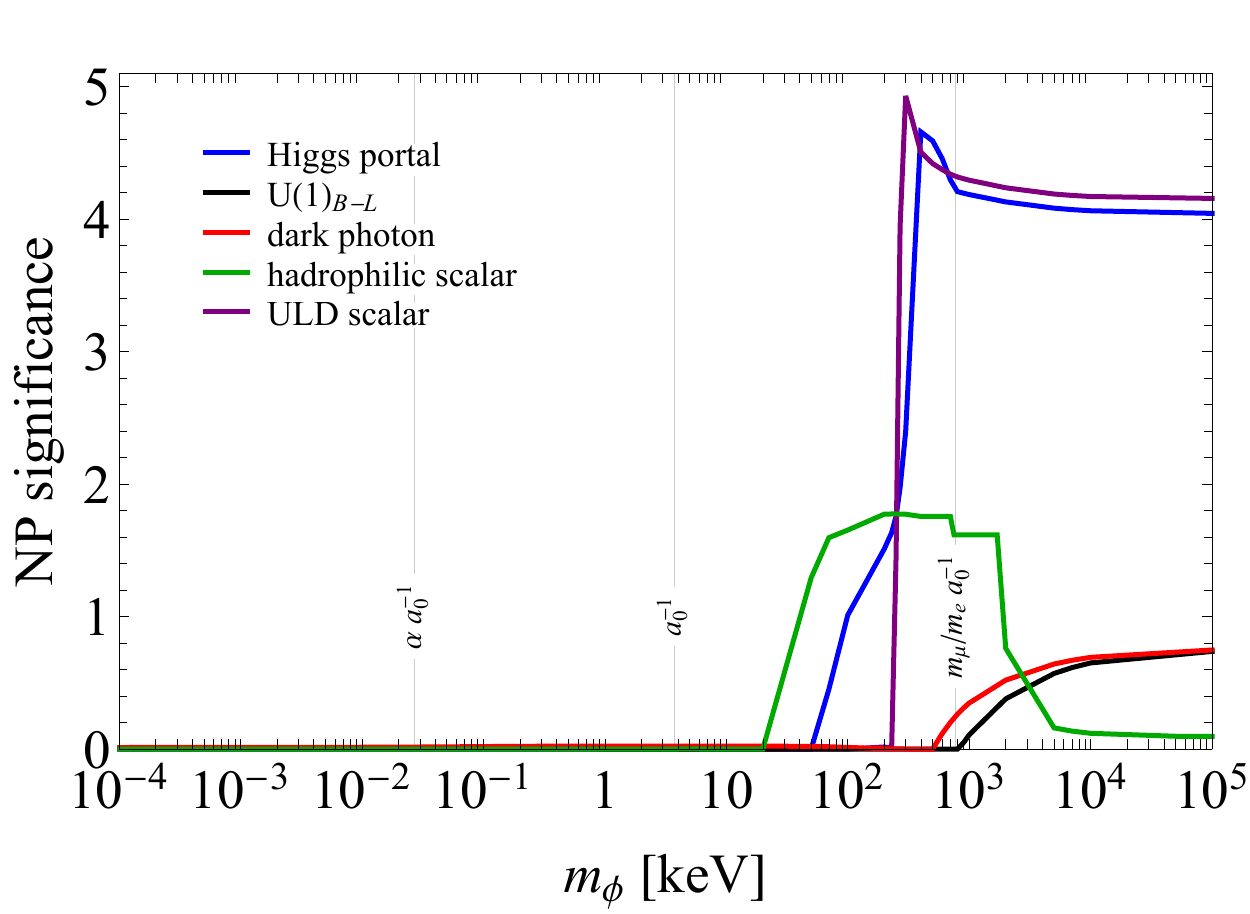}
    \caption{NP significance relative to the SM, $Z\equiv \sqrt{\chi^2_{\rm SM}-\chi^2_{\rm NP}}$, where $\chi^2_{\rm NP}$ ($\chi^2_{\rm SM}$) is the minimum $\chi^2$ with NP (without NP) for a fixed NP mass.}
    \label{fig:NPsigma}
\end{figure}

\begin{figure}
    \centering
    \begin{tabular}{cccc}
      \includegraphics{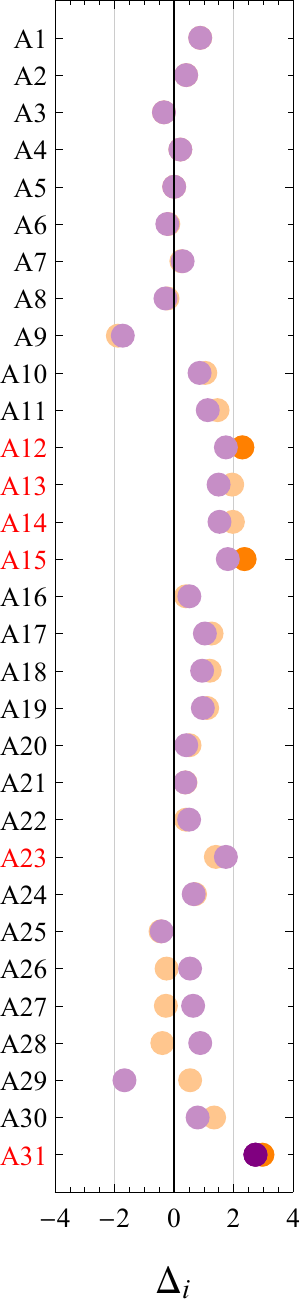}   &     \includegraphics{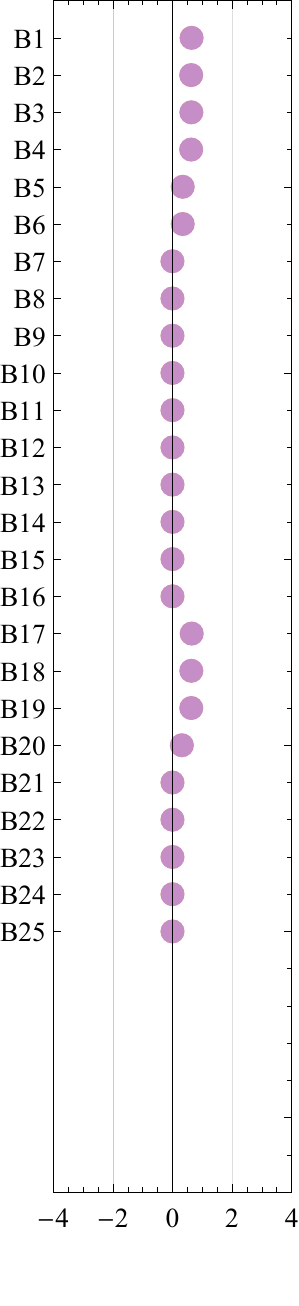} &    \includegraphics{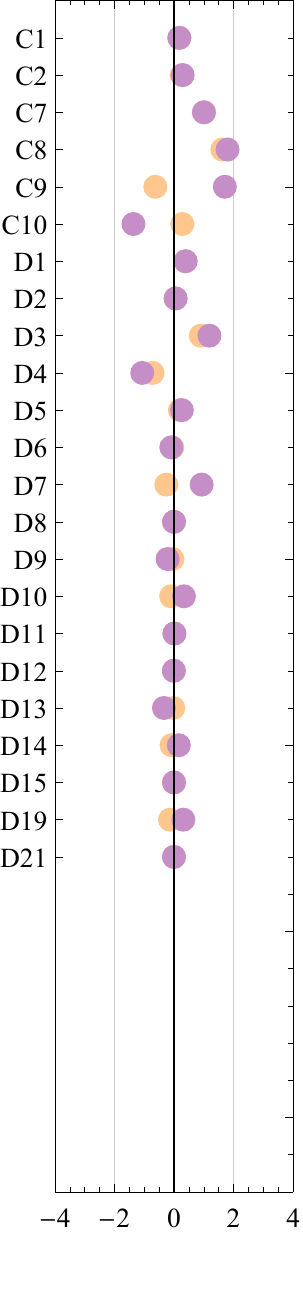} &    \includegraphics{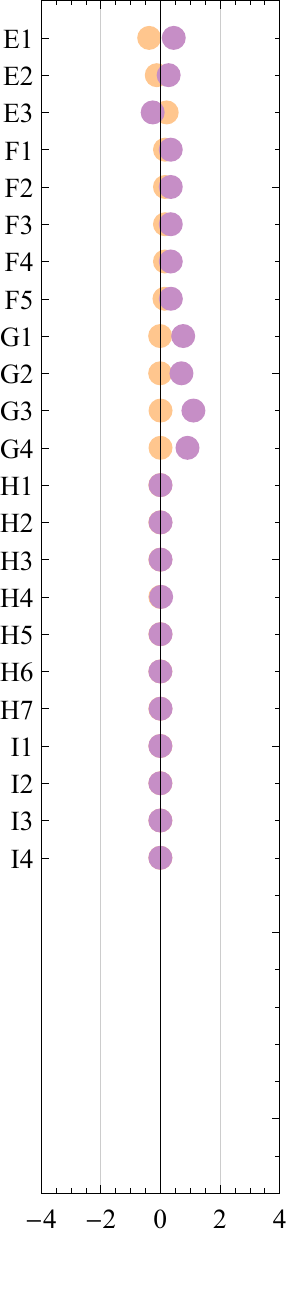}
   \end{tabular}
    \caption{Difference of squared normalized residuals, $\Delta_i\equiv {\rm sign}(\delta R_i^2)\sqrt{|\delta R_i^2|}$, of individual input data within the DATA22 dataset, where $\delta R_i^2\equiv (R_i^{\rm SM})^2-(R_i^{\rm NP})^2$ is the contribution of each input datum $i$ to the $\chi^2$ difference between SM and NP ignoring correlations with other input data. 
    $R_i^{\rm NP}$ are the normalized residuals of the scalar photon (orange) and the ULD scalar (purple) models, evaluated at their respective best-fit points. 
    Darker colors indicate input data with $\sqrt{|\delta R_i^2|}\geq 2$. 
    Input data satisfying $|R_i^{\rm SM}|\geq 2$ are indicated in red.}
    \label{fig:NPresidu}
\end{figure}

\begin{figure}
    \centering
     \includegraphics[width=0.45\columnwidth]{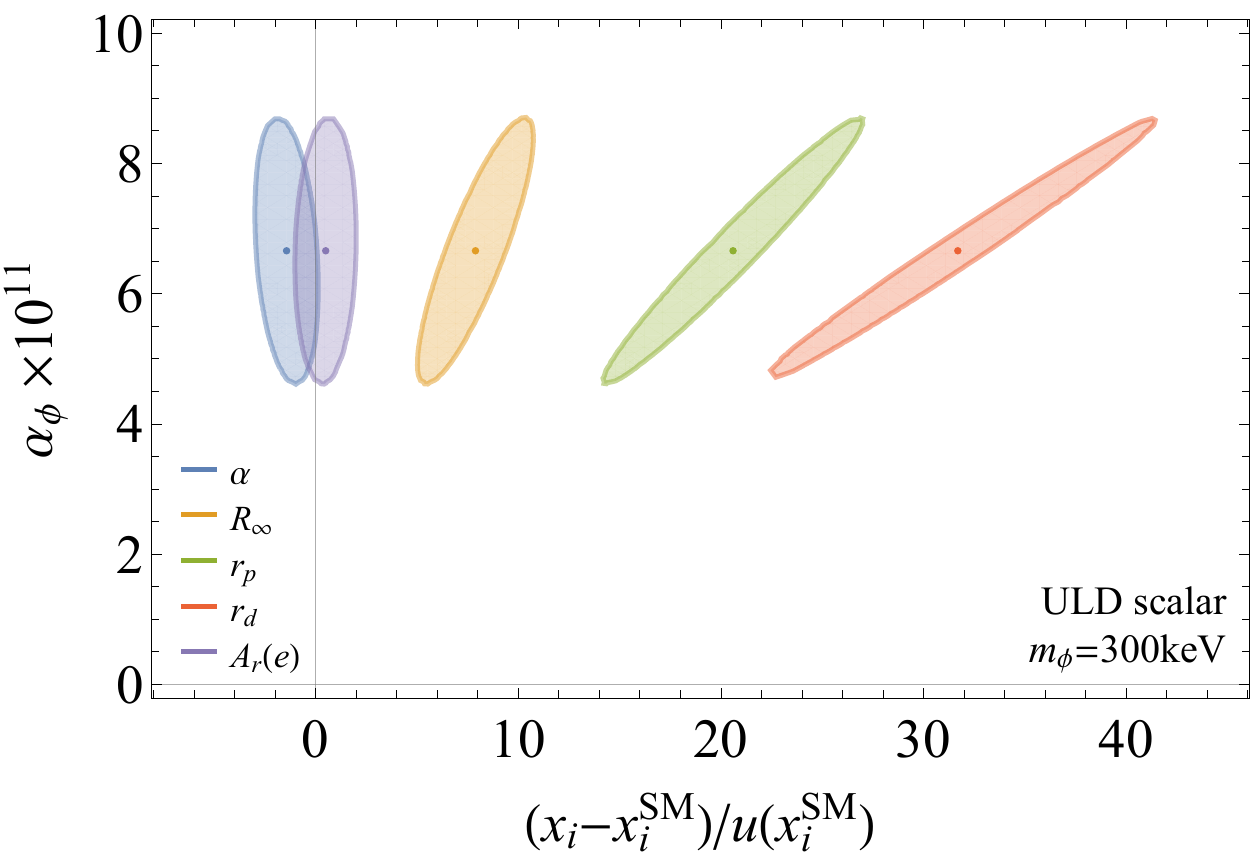}
    \caption{The 68\,$\%$~CL covariance ellipses of the
 fundamental constants $x_i=\alpha,R_\infty,r_p,r_d,A_{\rm r}(e)$ (color coding as indicated) and the NP coupling constant $\alphaNP$ for a $300\,$keV ULD scalar, using the DATA22 dataset. The $x$-axis is shifted by $x_i^{\rm SM}$, the extracted value of $x_i$ for the SM-only hypothesis, and normalized by its uncertainty.}
    \label{fig:alphaNP-vs-SM}
\end{figure}

The fit to DATA22 dataset assuming only the SM, {\it i.e.}, no NP,  gives the minimum $\chi^2$ per degree of freedom $\chi^2_{\rm SM}/\nu_{\rm dof}\simeq 1.4$ ($\nu_{\rm dof}=102-62=40$). Ignoring correlations, the largest  contributions to the $\chi^2$ are from the hydrogen observables A12--A15, A23 and A31 in Tables \ref{tab:inputsA} and \ref{tab:inputsNEW}, see also Fig.~\ref{fig:SMresidu}.

The adjustments exhibit no significant ($>2\sigma$) preference over the SM for the gauged $B-L$, dark photon and hadrophilic scalar models, see Fig.~\ref{fig:NPsigma}. The ULD scalar and the Higgs portal, on the other hand, show a preference over the SM for $m_\phi a_0\gtrsim m_\mu/m_e$. This NP evidence can already be anticipated from the bounds on $\alpha_\phi$ shown in Fig. \ref{fig:NPBound}. For masses heavier than $m_\phi a_0\gtrsim m_\mu/m_e$ the main constraint is from muonic hydrogen, so that one would expect, due to enhanced couplings of $\phi$ to muons, the bounds on ULD scalar and Higgs portal to be stronger by a factor $\cO( m_\mu/m_e)$ relative to all the other models we consider, in which $q_e=q_\mu$. The bounds in Fig. \ref{fig:NPBound}, however, are found  for the ULD and Higgs portal models to be weaker by a factor of $\cO(20)$ than the naive expectations, for heavy $\phi$.  
This is an indication that the data favors NP models with large $\mu$-to-$e$ coupling ratio over the SM in this mass range. 

The significance of the deviation is at the $\sim4\sigma$ level for the Higgs portal and the $\sim5\,\sigma$ level for the ULD scalar.   
Figure~\ref{fig:alphaNP-mNP} in the main text shows the preferred region in the ULD model parameters, where the best-fit point is $\mNP=300\,\keV$ and $\alphaNP=6.7\times 10^{-11}$. 
Similarly, the left panel in Fig.~\ref{fig:HP-SPexcess} shows the preferred region for the Higgs portal parameters, with the best-fit point given by $\mNP=400\,$keV and $\alphaNP=2.5\times 10^{-11}$. 
In both models the NP evidence  receives support mostly from the recent measurements of the hydrogen $2S_{1/2}-8D_{5/2}$ and $1S_{1/2}-3S_{1/2}$ transitions~\cite{Brandt:2021yor,Grinin:2020}, as well as muonic deuterium, see Fig.~\ref{fig:NPresidu}. 
These tensions between data and the SM prediction are not new. 
The authors of Ref.~\cite{Brandt:2021yor} already pointed out the inconsistency of their $2S_{1/2}-8D_{5/2}$ measurement with hydrogen theory and discussed NP interpretations in the form of Yukawa potential as well as its impact on the determination of $\Rinf$. 
The $1S_{1/2}-3S_{1/2}$ hydrogen and muonic deuterium Lamb shift measurements are known pieces of the so-called proton-radius puzzle~\cite{Karr:2020,Gao:2022}. 
Our analysis shows that both tensions can be significantly ameliorated by postulating the existence of a single light scalar mediator, with the pattern of couplings to the SM fermions such as in the Higgs portal or in the ULD model, with only the latter also avoiding other, non-spectroscopic constraints.

\section{Extraction of fundamental constants in the presence of NP}
\label{sec:app:extraction:NP}

In this section we discuss the effect of NP on the uncertainty of extracted fundamental constants.
Figure~\ref{fig:Correlation} in the main text shows the 68$\%$ CL region for simultaneous determinations of the proton charge radius $r_p$ and the Rydberg constant $R_\infty$,
assuming the SM-only hypothesis, using either the CODATA18 or DATA22 datasets, 
as well as for the ULD and the Higgs portal models (DATA22 only).
As expected, the extracted values of $r_p$ and $\Rinf$ are highly correlated regardless of the existence of NP.  NP induces significant shifts in the extracted values of $\Rinf$ and $r_p$, notably when the data shows evidence for nonzero $\alphaNP$.  
Furthermore, the larger the shift in the extracted value of the fundamental constant, the more important is the correlation with $\alphaNP$, see Fig.~\ref{fig:alphaNP-vs-SM}. 
 
\begin{figure}[t]
    \centering
    \begin{tabular}{cc}
    \includegraphics[width=0.45\columnwidth]{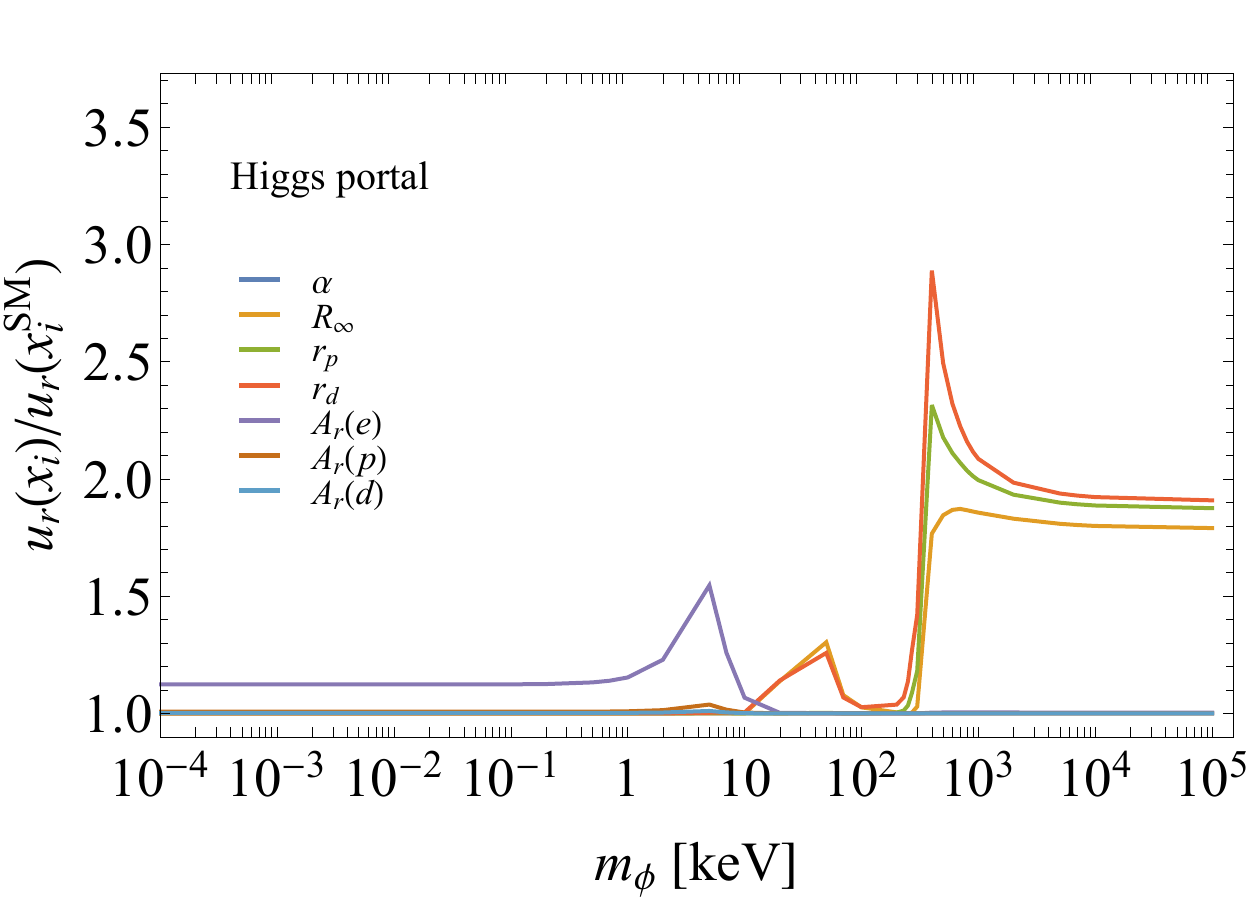}&
    \includegraphics[width=0.45\columnwidth]{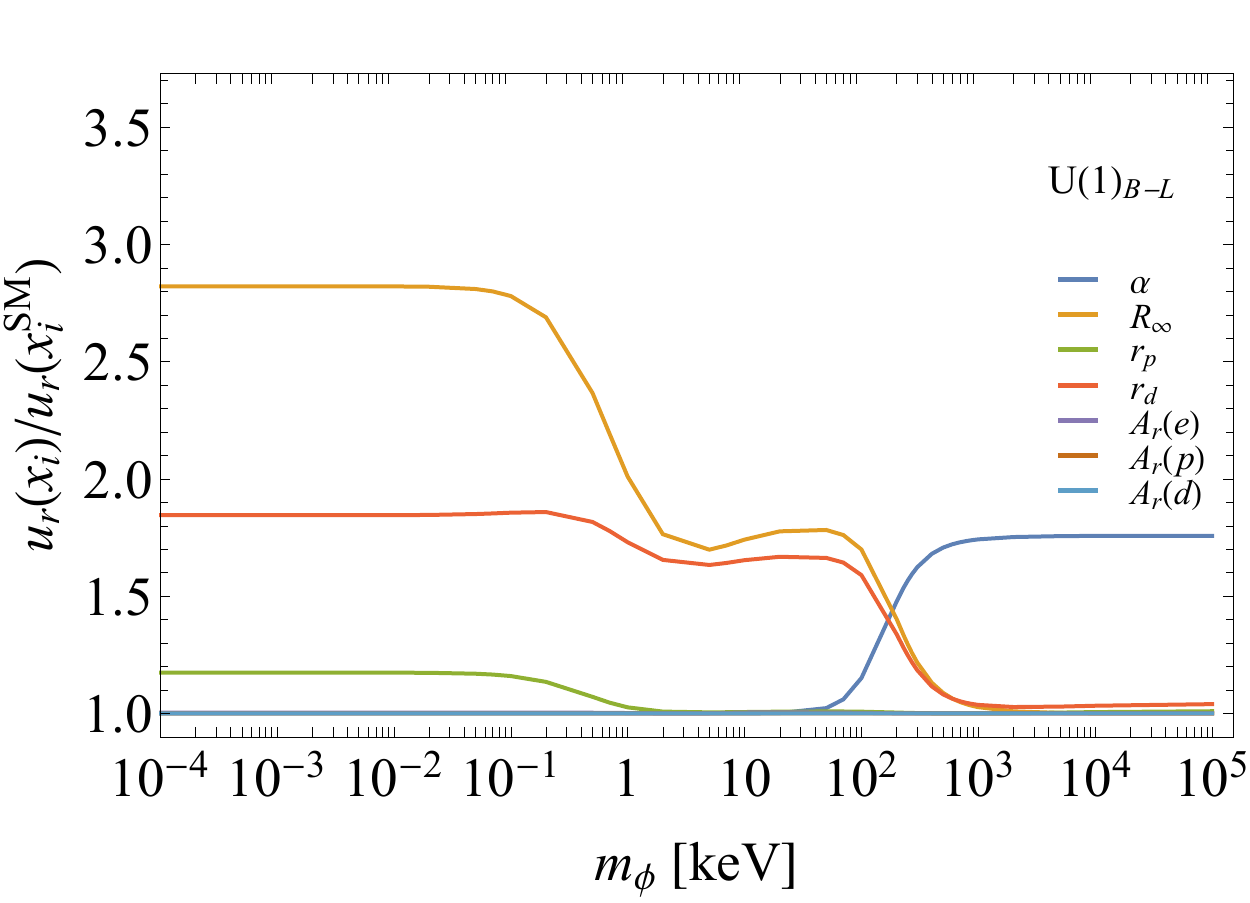}\\
    \includegraphics[width=0.45\columnwidth]{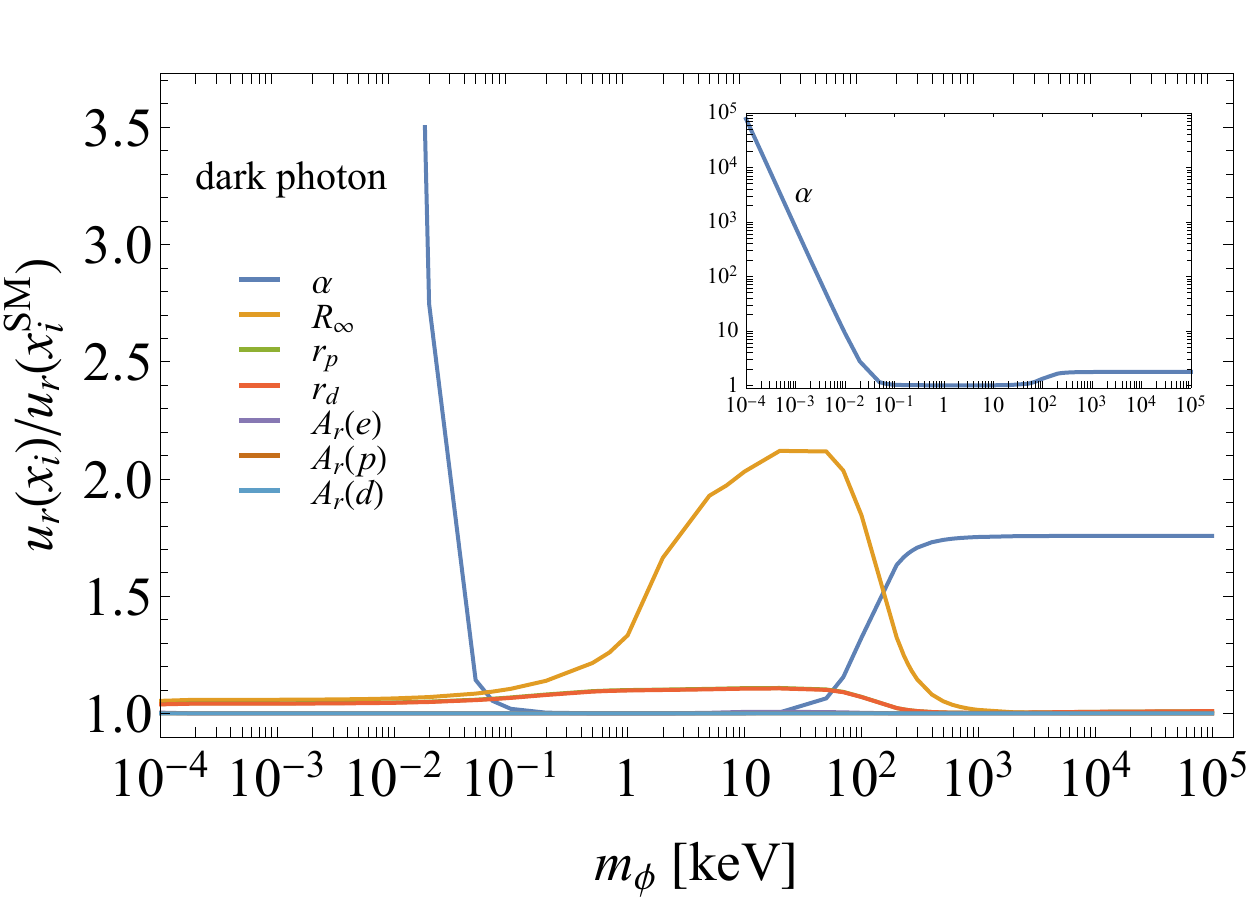}&  \includegraphics[width=0.45\columnwidth]{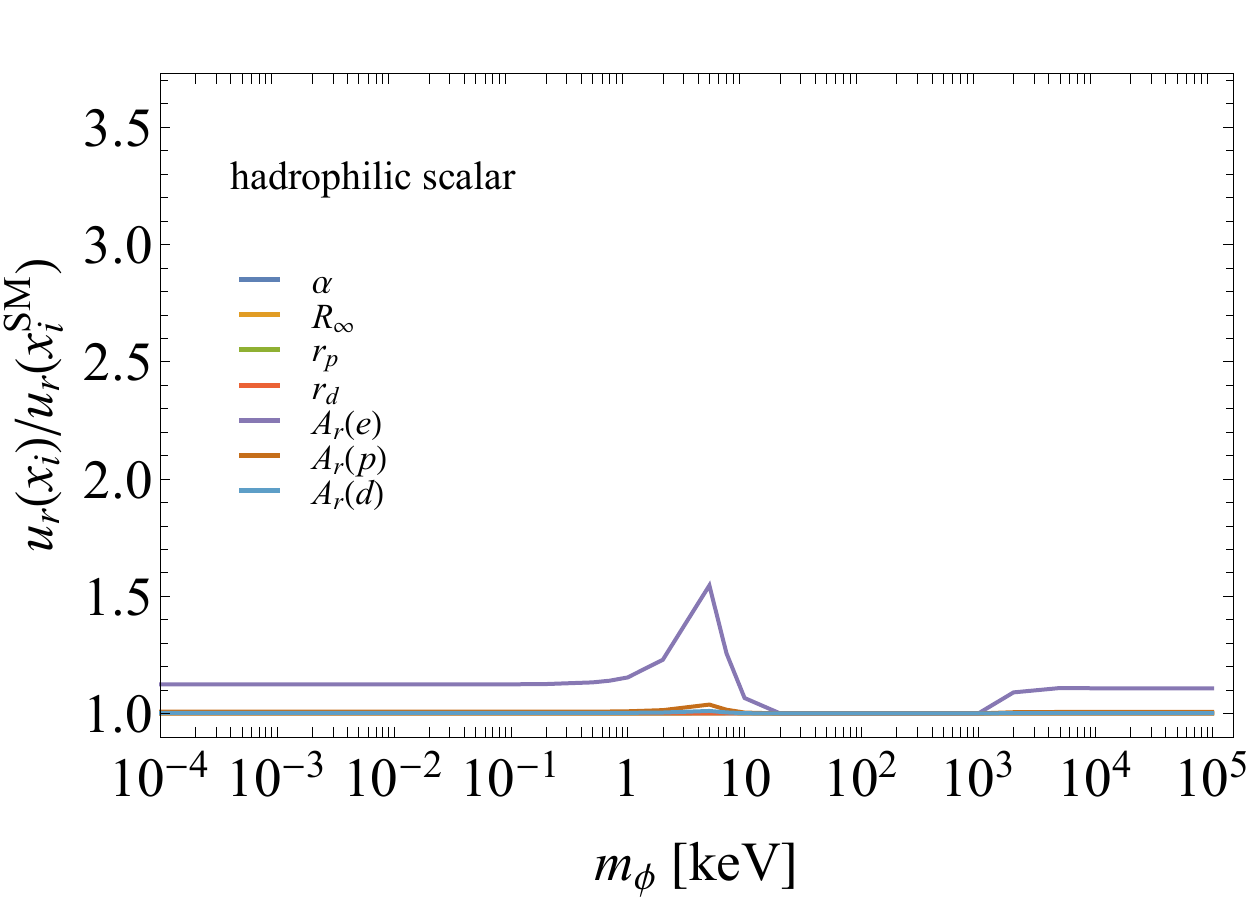}\\
    \includegraphics[width=0.45\columnwidth]{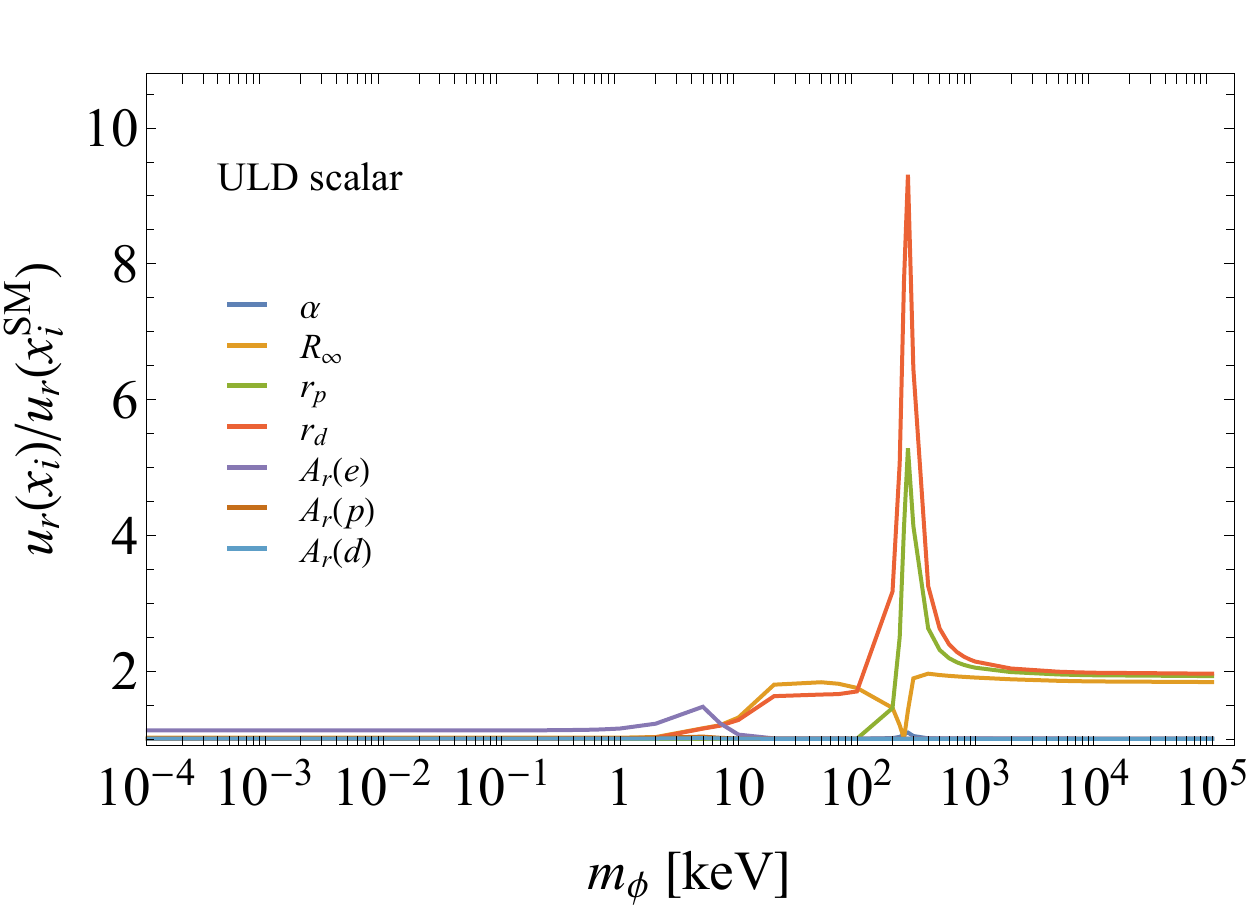}&
   \includegraphics[width=0.45\columnwidth]{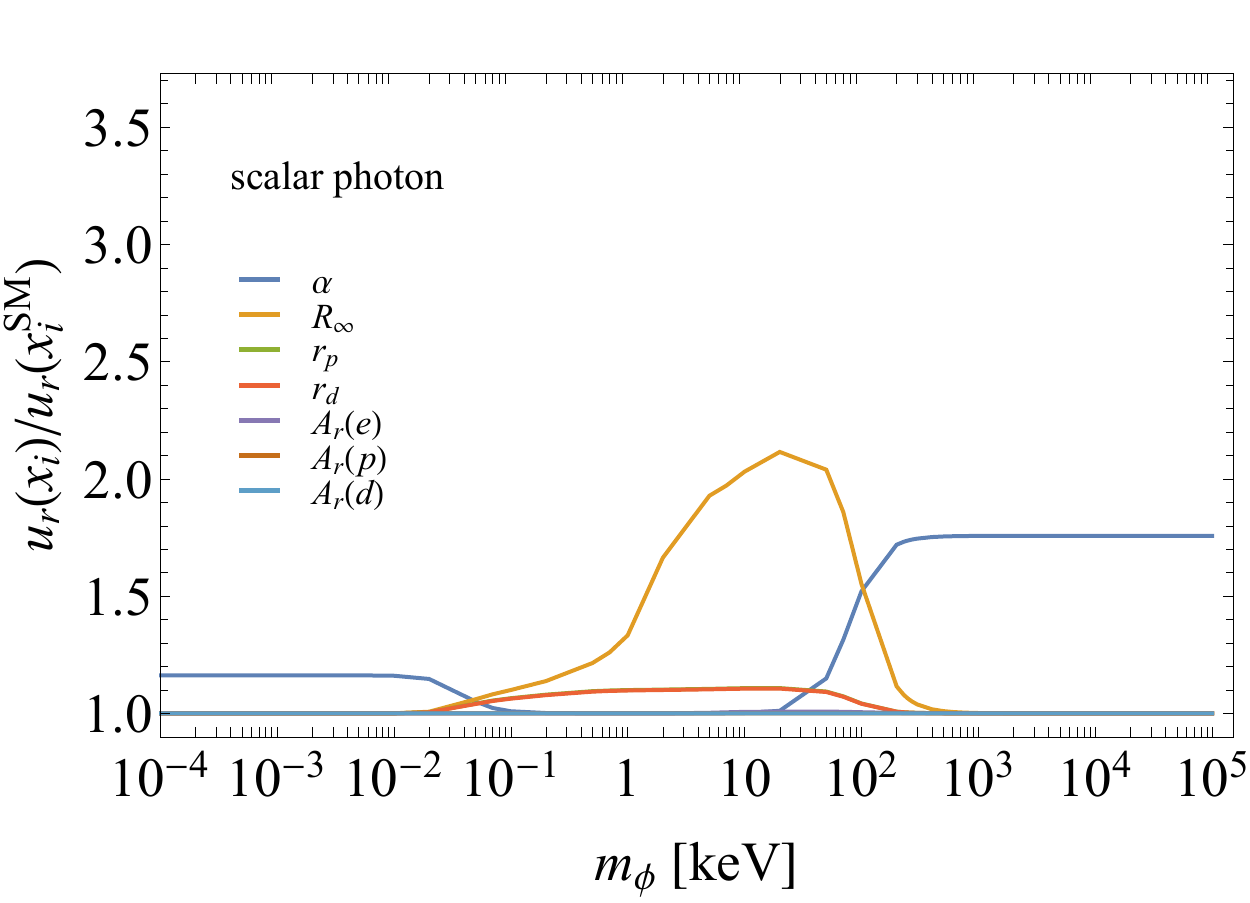}
    \end{tabular}
    \caption{The ratio of  relative uncertainties $u_r(x_i)\equiv u(x_i)/x_i$  on fundamental constants $x_i=\alpha,R_\infty,r_p,r_d,A_{\rm r}(e)$ (color coding as indicated) when extracted assuming NP relative to the extraction in the SM, for all six NP benchmark models.}
    \label{fig:uncertainties}
\end{figure}

Figure~\ref{fig:uncertainties} illustrates the impact of allowing in the fit the possibility of NP. The relative uncertainties on the extracted values of fundamental constants, $g_{\rm SM}$, are plotted as a function of the NP mass, for all six benchmark models. In many cases the uncertainties on the extracted fundamental constants change significantly, by factors of $\cO(1)$, even if the NP parameters are strongly disfavored by data.  
For dark photon, in particular, the uncertainty on $\alpha$ increases as $\sim \mNP^{-2}$ for $\phi$ masses below $\sim 10^{-2}\,$keV. This is a result of a degeneracy between  $\alpha$ and $\alphaNP$ in the $m_\phi\to 0$ limit, see  the inset in Fig.~\ref{fig:uncertainties}. The combination $\alpha+\alphaNP$, however,  is well determined.

\section{Further details on the $m_\phi\to0$ limit}
\label{sec:masslessNP}

The extraction of bounds on NP parameters in the $\mNP\to0$ limit requires special care, especially for new vector bosons that couple to the SM fermions in a similar way than the QED photon. Massless dark photon, in particular, is completely degenerate with the QED photon. 
In the massless limit all the effects due to the exchanges of a dark photon are therefore absorbed in the SM predictions by performing the shift $\alphaEM \to \alphaEM+\alphaNP$, {\it i.e.}, the massless dark photon is unobservable. The reason is that for a massless dark photon one can always choose a linear combination of photon and dark photon fields that does not couple to the SM currents, and redefine the orthogonal linear combination as the SM photon. This 
behavior should be reflected in theoretical predictions.
That is, in the $\mNP\to0$ limit, $\cO_{\rm SM}$ only depends on $\alphaEM+\alphaNP$, so that this sum can be interpreted as the SM fine-structure constant. Note that the shift $\alphaEM\to\alphaEM+\alphaNP $ also implies a redefinition of the Rydberg constant extracted from hydrogen, $\Rinf\to(\alphaEM+\alphaNP)^2m_e/(4\pi)$, and of the Bohr radius, $a_0^{-1}\to 4\pi \Rinf/(\alphaEM+\alphaNP)$. 

 The treatment of NP corrections in Eqs.~\eqref{eq:Vtilde}, \eqref{eq:Ovector} is  designed such that {\em a)} it is always correct to leading order in $\alpha_\phi$, and {\em b)} it reproduces the correct result for the massless dark photon, {\it i.e.}, for $m_\phi\to0$ and $q_i=Q_i$. The degeneracy of dark photon with the QED photon is  broken either by having $\mNP\neq 0$ (while still $q_i=Q_i$ for all particles) or by having a coupling for at least one particle differ from the QED one, $q_i\neq Q_i$.  For infinitesimal deformations from the massless dark photon limit the leading observable NP effect is captured by the potential $\widetilde{V}_\NP$. 
For  $\mNP a_0\ll1$, keeping the NP couplings still aligned with QED, $q_i=Q_i$, the leading  non-trivial term in $\tilde{V}_\NP$ scales as $\cO(\mNP^2)$, indicating a parametric loss of sensitivity to NP for small $m_\phi$, $\widetilde{\cO}_{\rm NP}\propto (\mNP a_0)^2$. (The term linear in $m_\phi$ is independent of $r$ and thus unobservable.) 
For massless vectors with couplings that deviate inifinitesimally from the QED couplings, {\it i.e.}, $q_i=Q_i+\delta q_i$ with $\delta q_i\ll Q_i$, the leading effect  scales as $\tilde{V}_\NP^{ij}\propto \delta q_i+ \delta q_j$. 

For massive vectors, $\mNP a_0\sim \cO(1)$, with charges that differ significantly from the QED ones, $q_i\ne Q_i$, the prescription in Eqs.~\eqref{eq:Vtilde}, \eqref{eq:Ovector} is equivalent to working to leading order in $V_{\rm NP}$ and not shifting the SM predictions at all. That is,  for $m_{\rm NP}$ nonzero and $q_i\ne Q_i$ such as massive $B-L$ gauge boson, both prescriptions are strictly speaking only correct to leading order in $\alpha_{\phi}$ and one can in principle use either of the two. However, for massless $B-L$ the prescription in Eqs.~\eqref{eq:Vtilde}, \eqref{eq:Ovector} will reproduce correctly many of the higher order corrections in hydrogen (though not all, for instance hadronic vacuum polarization type contributions are not correctly captured, as are higher order corrections for deuterium), and this may offer some benefit.

Finally, we point out that there is partial degeneracy also between QED photon and the massless scalar photon model with $q_i=Q_i$. However, the degeneracy of QED photon and massless scalar photon is only approximate. It is broken by relativistic corrections, which are different for vector and scalar degrees of freedom. This corrections are $\cO(\alpha^2)$ in atomic and molecular transitions. This means that it is consistent to work with unshifted SM predictions and add the NP contributions at leading order using unsubtracted potential $V_{\rm NP}$ also in the case of scalar photon, as we did for the other three scalar benchmarks in the main text.